\documentclass[review]{elsarticle}

\usepackage{lineno,hyperref}
\modulolinenumbers[5]

\journal{arxiv.org}

\usepackage[export]{adjustbox} 

\usepackage{color,soul}
\usepackage{setspace}

\usepackage{subfig} 
\usepackage{amsmath}
\usepackage[english]{babel}
\usepackage{lmodern}
\usepackage[T1]{fontenc}
\usepackage{algorithmic}
\usepackage{fancyvrb}
\usepackage{booktabs}
\usepackage{graphicx}
\usepackage{graphics}
\usepackage{caption}
\usepackage{float}

 \usepackage{amssymb}
\usepackage{palatino}
\usepackage{amsmath}
\usepackage{longtable}    
\usepackage{graphicx}
\usepackage{xcolor}
\usepackage{url}
\usepackage{multicol}
\usepackage{multirow}
\usepackage{array}
\usepackage{lscape}
\usepackage{boxedminipage}
\usepackage{multicol}
\usepackage{multirow}
\usepackage{color,soul}

\begin{document}

\begin{frontmatter}


\title{Reconstructed Discontinuous Galerkin Method for Compressible Flows in Arbitrary Lagrangian-Eulerian Formulation}

\author[]{Chuanjin Wang\corref{cor0}}
\ead{cwang35@ncsu.edu}
\author[]{Hong Luo\corref{cor1}}
\ead{hong\_luo@ncsu.edu}
\cortext[cor0]{Contribution from the first author was made when working at North Carolina State University.}

\address{
Department of Mechanical and Aerospace Engineering \\
North Carolina State University \\
Raleigh, NC 27695, USA}

\begin{abstract}

We present a high-order accurate reconstructed discontinuous Galerkin (rDG) method in arbitrary Lagrangian-Eulerian (ALE) formulation, 
for solving two-dimensional compressible flows on moving and deforming domains with unstructured curved meshes.
The Taylor basis functions in use are defined on the time-dependent domain, 
on which also the integration and computations are performed.
A third order ESDIRK3 scheme is employed for the temporal integration.
The Geometric Conservation Law (GCL) is satisfied by modifying the grid velocity terms 
on the right-hand side of the discretized equations at Gauss quadrature points.
To avoid excessive distortion and invalid elements near the moving boundary, 
we use the radial basis function (RBF) interpolation method for propagating the mesh motion 
of the boundary nodes to the interior of the mesh.
Several numerical examples are performed to verify the designed spatial and temporal orders of accuracy, 
and to demonstrate the ability of handling moving boundary problems of the rDG-ALE method.

\end{abstract}


\end{frontmatter}


\section{Introduction}
\label{sec:introduction}

Many engineering problems requires the solution on variable geometries, 
such as aeroelasticity,fluid-structure interaction, flapping flight 
and rotor-stator flows in turbine passage. 
Due to the limitation of the measurement techniques in the experiment, 
numerical simulation of such problems is an important supplement to 
investigating and understanding these complex phenomena. 
The arbitrary Lagrangian-Eulerian (ALE) \cite{hirt1974ale} formulation 
taking into account the mesh motion by nature, has been considered often 
and commonly used to solve such problems numerically. 

For these engineering problems mentioned above, the ALE formulation usually 
combines consistently the advection due to mesh motion and the advection 
due to fluid motion, i.e., these two are solved simultaneously.
This type of formulation is termed \textit{unsplit ALE} 
\cite{waltz2014aleFE3D,waltz2013aleSplitting,morgan2015alePCH}, 
as opposed to the \textit{split ALE} or \textit{Lagrange-plus-remap ALE} which consists of three steps, a Lagrangian step, a rezoning step and a remapping step. 
The Lagrangian-plus-remap ALE method is mainly used in hydrodynamics 
where the evolution of flow is undergoing large deformation due to the strong compression or expansion.
The application of the unsplit ALE method to the hydrodynamic problems has been reported in \cite{luo2004ale,waltz2014aleFE3D,morgan2015alePCH,morgan2017ale}.
In this paper, our focus is on the unsplit ALE method for the moving boundary/grid problems, typically the flow over moving airfoils.

In such problems, the velocity and/or position of the boundary nodes are usually known in practice, 
either from the prescription of an analytical expression or discrete data, 
or from the response of the solid in the fluid-structure interaction case. 
With the movement of the boundary nodes, the interior of mesh might be distorted or even invalid.
Thus, either regenerating the mesh, or a smoothing procedure that propagates 
the boundary motion into the interior is necessary, to preserve the mesh quality.
However, regeneration of the mesh is usually computationally very expensive, 
and the latter case, a smoothing procedure is preferred in general.
Various methods have been proposed in the literature to smooth the interior mesh. 
One type of these methods requires the solution of a system of 
elliptic (Poisson-type) partial differential equations (PDEs), such as linear elasticity \cite{mm:curve:hartmann} and non-linear elasticity \cite{mm:curve:persson} methods.
Another smoothing procedure is the interpolation method, for example, 
the radial basis function (RBF) interpolation \cite{mm:rbf,kashi2016aiaa,kashiThesis} and Delaunay graph mapping \cite{mm:dgm}.
In this work, we take advantage of the RBF interpolation based on its efficiency and mesh quality performance.

The high order methods, most commonly the discontinuous Galerkin methods (DGM), 
due to their higher computational efficiency compared with the low order method, 
have been investigated extensively in the Eulerian frame, 
and are gaining more and more interest in the ALE formulation.

The discontinuous Galerkin methods (DGM), originally introduced for solving the neutron 
transport equation by Reed and Hill \cite{reed1973triangularmesh}, are widely used in 
computational fluid dynamics (CFD), 
owing to their many distinctive, and attractive features, 
such as flexibility to handle complex geometry, 
compact stencil for higher-order reconstruction, and 
amenability to parallelization and \textit{hp}-adaptation. 
However, the DG methods have been recognized as expensive in terms of both computational 
costs and storage requirements. Indeed, compared to the finite element
and finite volume methods, the DGM require solutions of 
systems of equations with more unknowns for the same grids. 
It is our belief that a lack of efficiency is one of reasons 
that prevents the application of DGM to engineering-type problems.
In order to reduce high costs associated with the DGM,
Dumbser et al. \cite{dumbser2008unified,dumbser2009very,dumbser2010arbitrary} introduced 
a new family of reconstructed DGM, termed
P$n$P$m$ schemes and referred to as rDG (P$n$P$m$) in this paper, 
where P$n$ indicates that a piecewise polynomial of degree of $n$ is used 
to represent a DG solution, and P$m$ represents a reconstructed polynomial 
solution of degree of $m$ ($m\geqslant n$) that is used to compute the fluxes. 
The rDG(P$n$P$m$) schemes are designed to enhance the accuracy of the DGM 
by increasing the order of the underlying polynomial solution. The beauty of rDG(P$n$P$m$) 
schemes is that they provide a unified formulation for both FVM and DGM, 
and contain both classical FVM and standard DGM as two special cases of rDG(P$n$P$m$) schemes. 
When $n=0$, i.e. a piecewise constant polynomial is used to represent 
a numerical solution, rDG(P$0$P$m$) is nothing but classical high order finite volume schemes, 
where a polynomial solution of degree $m$ ($m\geqslant 1$) is reconstructed from 
a piecewise constant solution. When $m=n$, the reconstruction reduces to the identity operator, 
and rDG(P$n$P$n$) scheme yields a standard DG(P$n$) method. 
For $n>0$, and $n>m$, a new family of numerical methods from third-order of accuracy upwards 
is obtained.
A number of algorithms are 
proposed  \cite{van2007discontinuous,luo2010reconstructed, luo2011class,zhang2012class1,zhang2012class2,HrDGChengComputers} to construct the P$n$P$m$ schemes, and their spatial convergence and 
effectiveness are validated \cite{luo2011comparative,xia2014implicit,xia2014set, liu2017reconstructed,wang2017nka}.

Based on the higher performance of the rDG methods in the Eulerian frame, 
an extension to the arbitrary Lagrangian-Eulerian (ALE) formulation is natural and desirable. 
A number of DG-ALE methods in the literature have been proposed and investigated for the compressible flows.
How to satisfy the Geometric Conservation Law (GCL) is a critical issue for the ALE formulation, 
especially for higher-order DG methods, 
where the basis functions being defined on the time-dependent or fixed reference domain/element will come into the picture, 
making the problem more complicated.

In general, there are two types of approaches to ensure the satisfaction of the GCL condition.
The most consistent and elegant approach is the space-time DG formulation.
This formulation is fully conservative in space and time, and the GCL is automatically satisfied.
The space-time DG methods have been applied to both incompressible \cite{van2007spaceTimeIncomp,cockburn2012spaceTimeIncomp} 
and compressible flows \cite{van2002spaceTimeI,van2002spaceTimeII, klaij2006spaceTime, wang2015spaceTime}.
However, the generation of the space-time meshes will require additional work.
And, the space-time DG methods do not allow for explicit time stepping or implicit multi-step schemes \cite{persson2009ale}.

The second type requires special treatment for the ALE formulations on a fixed or time-dependent mesh, 
for example, an additional equation for updating the transformation Jacobian, or a correction of the grid velocity terms.
Lomtev et al. \cite{lomtev1999ale} presented a matrix-free DG-ALE
method using spectral basis for 2D and 3D compressible viscous flows in moving domains. 
A force-directed algorithm from graph theory is used to update the grid while minimizing distortions.
In the method proposed by Persson et al. \cite{persson2009ale}, 
the governing equations in the time-dependent physical domain are transformed to 
the conservations laws in a fixed reference domain (the initial domain), 
the conservative variables defined on the initial domain are solved thereafter. 
A continuous mapping between the time-dependent physical domain and the fixed 
initial domain is introduced to take into account the mesh motion. 
The transformation Jacobian is updated in time to ensure the GCL condition.
Nguyen \cite{nguyen2010ale} presented a DG-ALE method using an
explicit fourth order TVD Runge-Kutta method and the freestream
solution are shown to be preserved numerically.
Mavriplis \cite{mavriplis2011ale} derived a general approach inspired 
from the space-time formulation to update the grid velocity terms at 
the Gauss quadrature points such that the GCL is satisfied.
Ren and Xu et al. \cite{renxu2016ale} introduced a DG-ALE method based 
on the gas-kinetic theory. 
In their method, the initial domain is taken as the reference domain, 
and the basis functions are mapped from this reference domain to the 
time-dependent physical domain. 
The computations are conducted in the physical domain. 
A space-time type integration is used to obtain the discretized 
equations and then the gas kinetic flux is computed to advance the solution. 
This method is shown to preserve the uniform flow automatically, and 
applied to several moving boundary problems.

In the current work, the reconstructed discontinuous Galerkin (rDG) 
method is extended to the ALE formulation, to simulate 
flow problems with moving or deforming grids.
The Taylor basis functions are defined on the time-dependent physical 
domain, as a continuation of the rDG method in the Eulerian formulation. 
For better resolution at the wall of the airfoil, the curved elements are used in the whole domain.
The third order temporal scheme ESDIRK3 (Explicit first stage, Single 
Diagonal coefficient, diagonally Implicit Runge-Kutta) is employed for time marching. 
To enforce the GCL, we follow a similar idea as in 
\cite{mavriplis2011ale}, by updating the grid velocity terms at Gauss quadrature points.
The radial basis function (RBF) interpolation method has been chosen 
as the mesh smoothing algorithm to provide the mesh motion for the 
interior nodes, given the displacement or velocity at the boundary nodes.
Numerical test cases are set up to demonstrate the accuracy and 
capability of the derived rDG-ALE method.

The remainder of this paper is organized as follows.
The governing equations will follow next.
The rDG formulation in ALE frame is derived in Section  \ref{sec:RDGALE}. 
Section \ref{sec:GCL} discusses how to satisfy the Geometric Conservation Law (GCL) condition.
The RBF method for the mesh movement will be described in Section \ref{sec:meshMotion}.
Section \ref{sec:numerical-examples} presents a set of numerical examples. 
Finally conclusions are given in Section \ref{sec:conclusion}.

\section{Governing Equations}

  The Navier-Stokes equations governing the unsteady compressible viscous flows
can be expressed as

\begin{equation}
  \label{euler:eq:governing-equations}
  \frac{\partial\textbf{U}}{\partial t}
  +\frac{\partial\textbf{F}_k(\textbf{U})}{\partial x_k}
  =\frac{\partial\textbf{G}_k(\textbf{U},\nabla \textbf{U})}{\partial x_k}
\end{equation}
where the summation convention has been used. The conservative variable $\textbf{U}$, advective flux vector $\textbf{F}$ and viscous flux vector $\textbf{G}$ are defined by

\begin{equation}
  \label{euler:eq:conservative-variables}
  \textbf{U}=
  \begin{pmatrix} 
    \rho \\ \rho u_i \\ \rho e 
  \end{pmatrix}
  \hspace{0.4in}
  \textbf{F}_j=
  \begin{pmatrix} 
    \rho u_j \\ \rho u_i u_j + p\delta_{ij} \\ u_j(\rho e +p) 
  \end{pmatrix}
  \hspace{0.4in}
  \textbf{G}_j=
  \begin{pmatrix} 
    0 \\ \tau_{ij} \\ u_l\tau_{lj}+q_j 
  \end{pmatrix}
\end{equation}

Here $\rho$, $p$ and $e$ denote the density, pressure and specific 
total energy of the fluid, respectively, and $u_i$ is the velocity 
component of the flow in the coordinate direction $x_i$. The pressure 
can be computed from the equation of state

\begin{equation}
  p=(\gamma -1)\rho\left(e-\frac{1}{2}u_i u_i\right)
\end{equation}
which is valid for perfect gas. The ratio of the specific heats $\gamma$ is assumed to be constant and equal to 1.4. 
The viscous stress tensor $\tau_{ij}$ and heat flux vector $q_j$ are given by

\begin{equation}
  \tau_{ij} = 
  \mu\left(
  \frac{\partial u_i}{\partial x_j} 
  +\frac{\partial u_j}{\partial x_i}
  \right)
  -\frac{2}{3}\mu\frac{\partial u_k}{\partial x_k}\delta_{ij}
  \hspace{0.4in}
  q_j = \frac{1}{\gamma -1}\frac{\mu}{Pr}\frac{\partial T}{\partial x_j}
\end{equation}
In the above equations, $T$ is the temperature of the fluid, $Pr$ the 
laminar Prandtl number, which is taken as 0.7 for air. $\mu$ 
represents the molecular viscosity, which can be determined through Sutherlands law

\begin{equation}
  \frac{\mu}{\mu_0}=\left(\frac{T}{T_0}\right)^{\frac{3}{2}} \frac{T_0+S}{T+S}
\end{equation}
where $\mu_0$ is the viscosity at the reference temperature $T_0$ and $S=110K$. The temperature $T$ of the fluid is determined by

\begin{equation}
  T = \frac{p}{\rho R}
\end{equation}
where $R$ is the gas constant.

The Euler equations can be obtained if the effects of viscosity and 
thermal conduction are neglected in Eq. \ref{euler:eq:governing-equations}.

\section{Reconstructed Discontinuous Galerkin Discretization}
\label{sec:RDGALE}
Since the grid velocity only contributes to and appears in the 
convective term, it's sufficient to consider the compressible Euler 
equations here for deriving the ALE formulation.
The differential governing equation can be written in the form
\begin{equation}
  \frac{\partial\textbf{U}(\textbf{x},t)}{\partial t} 
  +\nabla \cdot \textbf{F(\textbf{U})}
= 0
\end{equation}
where $\frac{\partial}{\partial t}$ is the usual Eulerian time derivative.

Defining the (Taylor) basis functions $\phi_j$ associated with the 
moving volume $\Omega_e^t$, multiplying $\phi_j$ on the equation above 
and integrating on the moving volume, one will get immediately 
\begin{equation}
  \label{eq:integral:eulerian}
  \int_{\Omega_e^t}\phi_j\frac{\partial\textbf{U}}{\partial t} d\Omega 
  + \int_{\Omega_e^t}\phi_j  \nabla\cdot\textbf{F} d\Omega
  =0
\end{equation}

Using Reynolds Transport Theorem
\begin{equation}
\begin{split}
\begin{aligned}
  \frac{d}{dt}\int_{\Omega_e^t}\textbf{U}\phi_j d\Omega
  &=
  \int_{\Omega_e^t}\frac{\partial (\textbf{U}\phi_j)}{\partial t} d\Omega 
  +
  \int_{\Omega_e^t}\nabla\cdot(\textbf{U}\phi_j \textbf{V}_g)d\Omega\\
  &=
  \int_{\Omega_e^t}\phi_j\frac{\partial\textbf{U}}{\partial t}  d\Omega
  +
  \int_{\Omega_e^t}\textbf{U}\frac{\partial \phi_j}{\partial t}   d\Omega
  +
  \int_{\Gamma_e^t}\textbf{U}\phi_j \textbf{V}_g\cdot\textbf{n}  d\Gamma
\end{aligned}
\end{split}
\end{equation}
where $\textbf{V}_g$ is the grid velocity, and the divergence theorem
\begin{equation}
  \int_{\Omega_e^t}\phi_j  \nabla\cdot\textbf{F} d\Omega
  +\int_{\Omega_e^t}\textbf{F}\cdot\nabla\phi_j   d\Omega
  =
  \int_{\Gamma_e^t}\phi_j \textbf{F}\cdot\textbf{n}d\Gamma
\end{equation}
Eq. \ref{eq:integral:eulerian} becomes
\begin{equation}
  \label{eq:rdg-ale:0}
  \frac{d}{dt}\int_{\Omega_e^t}\textbf{U}\phi_j d\Omega
  +
  \int_{\Gamma_e^t}\phi_j  (\textbf{F}-\textbf{U}\textbf{V}_g)\cdot \textbf{n} d\Gamma
  -
  \int_{\Omega_e^t} (\textbf{F}\cdot\nabla\phi_j + \textbf{U}\frac{\partial \phi_j}{\partial t} ) d\Omega
  = 0
\end{equation}
The fundamental ALE relation for the total time derivative, Eulerian 
time derivative and the spatial gradient is
\begin{equation}
  \frac{d\phi}{d t}
  =
  \frac{\partial\phi}{\partial t}
  +\textbf{V}_g\cdot  \nabla\phi
\end{equation}
We note that the total time derivative $\frac{d}{d t}$ is with respect 
to the grid velocity, in contrast to the material derivative with 
respect to the fluid velocity.
And also, the $\frac{d}{d t}$ here is equivalent to the referential 
time derivative $\frac{\partial{}}{\partial t}|_X$ in 
\cite{renxu2016ale} where the subscript $X$ was used to indicate 
fixing the mapping position in the initial mesh; in our case, although 
we are not performing an explicit mapping from the initial 
configuration to the time-dependent element as in \cite{renxu2016ale}, 
we are essentially using an implicit one-to-one mapping from the 
points (e.g., the Gauss quadratures) on a standard reference element 
to the grid coordinates in the time-dependent element. 
Thus in either case, we are tracking the grids when using this total 
time derivative, i.e., holding the grid point label fixed.

Plugging this relation into Eq. \ref{eq:rdg-ale:0}, we end up with the rDG-ALE formulation
\begin{equation}
  \label{eq:rdg-ale}
  \frac{d}{dt}\int_{\Omega_e^t}\textbf{U}\phi_j d\Omega
  +
  \int_{\Gamma_e^t}\phi_j  (\textbf{F}-\textbf{U}\textbf{V}_g)\cdot \textbf{n} d\Gamma
    -
  \int_{\Omega_e^t} (\textbf{F}-\textbf{U}\textbf{V}_g) \cdot\nabla\phi_j d\Omega
  -
  \int_{\Omega_e^t}\textbf{U}\frac{d \phi_j}{d t} d\Omega
  = 0
\end{equation}
It's worth noting that our basis functions are defined on the time-
dependent physical element, thus the basis functions are also time-
dependent, i.e.,
$$\phi_j = \phi_j(t)$$
except for $j=1$, $\phi_1 = 1$ which is a constant and thus ${d \phi_1}/{d t}$ is zero everywhere and at any instant time. 
The resulting ALE formulation will be discretized in space using the 
reconstructed discontinuous Galerkin method as in the Eulerian 
formulation \cite{luo2010reconstructed,luo2011class, luo2011comparative,HrDGChengComputers,wang2017nka}.

\section{Geometric Conservation Law (GCL)}
\label{sec:GCL}
The Geometric Conservation Law (GCL) states that the fully discretized equations
should preserve a constant solution under arbitrary mesh movement, i.e., given
a uniform initial flow condition, the solution should be maintained by the devised ALE solver.

Consider the BDF1 temporal discretization scheme. 
\begin{equation}
\label{eq:GCL:BDF1:U}
\begin{split}
\begin{aligned}
  \frac{1}{\Delta t}\left[
  \int_{\Omega_e^{n+1}}\textbf{U}^{n+1}\phi_j^{n+1} d\Omega
  -\int_{\Omega_e^n}\textbf{U}^n\phi_j^n d\Omega
  \right]
  +
  \int_{\Gamma_e^{n+1}}\phi_j^{n+1}  (\textbf{F}^{n+1}-\textbf{U}^{n+1}\textbf{V}_g^{n+1})\cdot \textbf{n} d\Gamma
  \\
  -
  \int_{\Omega_e^{n+1}} (\textbf{F}^{n+1}-\textbf{U}^{n+1}\textbf{V}_g^{n+1}) \cdot\nabla\phi_j^{n+1} d\Omega
  -
  \int_{\Omega_e^{n+1}}\textbf{U}^{n+1}\frac{d \phi_j^{n+1}}{d t}   d\Omega
  = 0
\end{aligned}
\end{split}
\end{equation}
Plugging a constant solution into the equation above leads to
\begin{equation}
\label{eq:GCL:BDF1:gcl}
\begin{split}
\begin{aligned}
  \frac{1}{\Delta t}\left[
  \int_{\Omega_e^{n+1}}\phi_j^{n+1} d\Omega
  -\int_{\Omega_e^n}\phi_j^n d\Omega
  \right]
  =
  \int_{\Gamma_e^{n+1}}\phi_j^{n+1} \textbf{V}_g^{n+1}\cdot \textbf{n} d\Gamma
  \\
  -
  \int_{\Omega_e^{n+1}} \textbf{V}_g^{n+1} \cdot\nabla\phi_j^{n+1} d\Omega
  +
  \int_{\Omega_e^{n+1}}\frac{d \phi_j^{n+1}}{d t}   d\Omega
\end{aligned}
\end{split}
\end{equation}
This is the GCL equation that needs to be satisfied by the rDG method with BDF1 scheme.
Note that for a constant solution $\textbf{U}$, the Eulerian flux $\textbf{F}$ vanishes due to its consistency property.
Unfortunately, this equation will not hold in general, even at the 
DG(P0) (Finite Volume) level which corresponds to $\phi = 1$. 
To verify, plugging in the basis $\phi = 1$, the GCL equation reduces to
\begin{equation}
\begin{split}
\begin{aligned}
  \frac{1}{\Delta t}\left[
  \Omega_e^{n+1} - \Omega_e^n
  \right]
  =
  \int_{\Gamma_e^{n+1}} \textbf{V}_g\cdot \textbf{n} d\Gamma
\end{aligned}
\end{split}
\end{equation}
where the mesh velocity is defined as
\begin{equation}
  \textbf{V}_g = \frac{\textbf{x}^{n+1} - \textbf{x}^n}{\Delta t}
\end{equation}
The GCL equation above for FV states that the volume change between two successively 
discretized time levels $n$ and $n+1$, equals to the volume flux at the element interfaces
at time level $n+1$ (time level $n$ if using explicit time stepping). Unfortunately, this is not
true in general.

Here we follow the idea from the work of Mavriplis et al. \cite{mavriplis2011ale}.
The major difference between the current work and that in  \cite{mavriplis2011ale} is that,
the basis functions in  \cite{mavriplis2011ale} are defined on the reference element, while
the Taylor-basis functions used in the current work are defined on the time-dependent physical element. 
This difference leads to two consequences. 
First, in \cite{mavriplis2011ale}, the substantial derivative of the 
basis functions with respect to the mesh motion will vanish, due to the definition of
the basis function and the fact that the reference element is invariant in time-- thus the basis functions
simply move with the grid velocity and their values will never change. In contrast, for the Taylor-basis 
in this work, they are defined on the time-dependent element, and if we consider a quadrature point, although
it also travels with the mesh velocity, the value of its basis function is changing in time inherently.
Thus, an extra term containing the total derivative of the basis functions will appear in this formulation.
Second, in \cite{mavriplis2011ale}, the solution expansion is 
performed on the reference element using the basis functions therein, 
thus the domain integral takes on a different form than this work, 
i.e., in \cite{mavriplis2011ale} the inverse of the transformation 
Jacobian emerges when transforming the spatial gradient from reference 
element to physical element, while in the current work the spatial 
gradient on the physical element can be used directly without any transformation.

The idea of how to enforce the GCL condition stems from the space-time DG method.
Performing a full integration on Eq. \ref{eq:rdg-ale} in both space and time (--on a space-time element) leads to
\begin{equation}
\begin{split}
\begin{aligned}
  \int_{\Omega_e^{n+1}}\textbf{U}^{n+1}\phi_j^{n+1} d\Omega
  -\int_{\Omega_e^n}\textbf{U}^n\phi_j^n d\Omega
  +
  \int_t^{t+\Delta t}
  \int_{\Gamma_e^t}\phi_j  (\textbf{F}-\textbf{U}\textbf{V}_g)\cdot \textbf{n} d\Gamma dt
  \\
  -
  \int_t^{t+\Delta t}
  \int_{\Omega_e^t} (\textbf{F}-\textbf{U}\textbf{V}_g) \cdot\nabla\phi_j d\Omega dt
  -
  \int_t^{t+\Delta t}
  \int_{\Omega_e^t}\textbf{U}\frac{d \phi_j}{d t}   d\Omega dt
  = 0
\end{aligned}
\end{split}
\end{equation}

\begin{figure}[H]
  \centering
  \subfloat{\includegraphics[trim= 6.3cm 8.5cm 8.3cm 2cm,clip,height = 2in]  
  {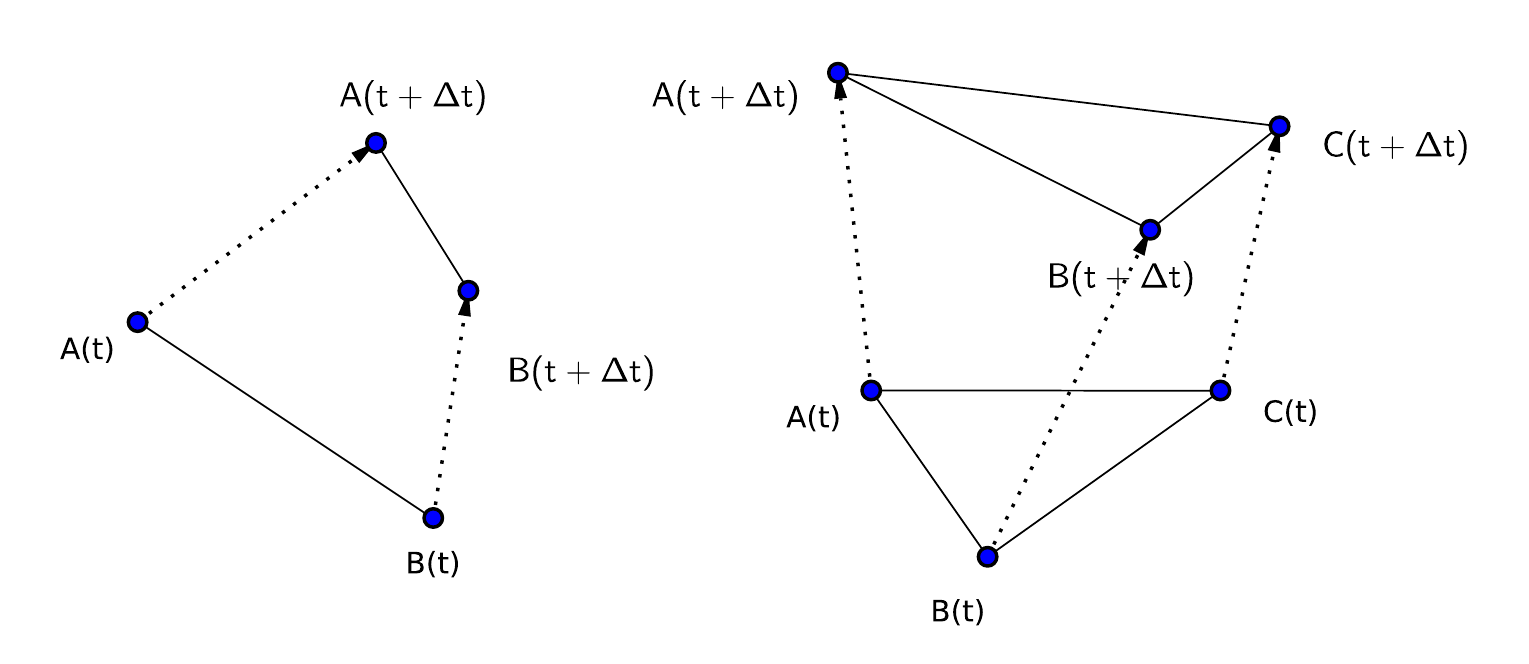}}  
  \caption{Schematics for space-time integration at an edge (left) and cell (right).}
  \label{fig:spacetime}
\end{figure}

The GCL equation for space-time DG method is then obtained by plugging in a constant solution
\begin{equation}
\label{eq:GCL:spaceTime:gcl}
\begin{split}
\begin{aligned}
  \int_{\Omega_e^{n+1}}\phi_j^{n+1} d\Omega
  -\int_{\Omega_e^n}\phi_j^n d\Omega
  =
  \int_t^{t+\Delta t}
  \int_{\Gamma_e^t}\phi_j  \textbf{V}_g\cdot \textbf{n} d\Gamma dt
  \hspace{1.0in}\\
  -
  \int_t^{t+\Delta t}
  \int_{\Omega_e^t} \textbf{V}_g \cdot\nabla\phi_j d\Omega dt
  +
  \int_t^{t+\Delta t}
  \int_{\Omega_e^t}\frac{d \phi_j}{d t}   d\Omega dt
\end{aligned}
\end{split}
\end{equation}
This GCL equation is automatically satisfied, provided enough Gauss 
quadrature points are employed in both space and time to conduct the numerical integration.
 
Inspired from this property of space-time integration, a 
straightforward way to satisfy the BDF1 GCL equation is to modify the 
right-hand side of Eq. \ref{eq:GCL:BDF1:gcl}  such that it works the 
same way as its space-time DG counterpart Eq. \ref{eq:GCL:spaceTime:gcl}.
More explicitly, we want to have the following conditions hold
\begin{equation}
\begin{split}
\left\{
\begin{aligned}
  \int_{\Gamma_e^{n+1}}\phi_j^{n+1} \textbf{V}_g^{n+1}\cdot \textbf{n} d\Gamma
  &=
  \frac{1}{\Delta t} \int_t^{t+\Delta t}
  \int_{\Gamma_e^t}\phi_j  \textbf{V}_g\cdot \textbf{n} d\Gamma dt
  \\
  \int_{\Omega_e^{n+1}} \textbf{V}_g^{n+1} \cdot\nabla\phi_j^{n+1} d\Omega
  &=
  \frac{1}{\Delta t} \int_t^{t+\Delta t}
  \int_{\Omega_e^t} \textbf{V}_g \cdot\nabla\phi_j d\Omega dt
  \\
  \int_{\Omega_e^{n+1}}\frac{d \phi_j^{n+1}}{d t}   d\Omega
  &=
  \frac{1}{\Delta t} \int_t^{t+\Delta t}
  \int_{\Omega_e^t}\frac{d \phi_j}{d t}   d\Omega dt
\end{aligned}
\right.
\end{split}
\end{equation}

However, as addressed in \cite{mavriplis2011ale}, although the values 
for these spatial integrals of grid velocities can be obtained, they 
do not yield grid velocities themselves, and hence can not be directly 
used in evaluating the integrals in Eq. \ref{eq:GCL:BDF1:U}.

Recall that eventually the numerical integration will be performed on 
the reference element where the Gauss quadrature points reside, to 
this end, it's beneficial to write the above equations in terms of 
numerical integration on the reference element.  

For BDF1 equation
\begin{equation}
\begin{split}
\begin{aligned}
  \frac{1}{\Delta t}\left[
  \int_{\widehat{\Omega}_e}\textbf{U}^{n+1}\phi_j^{n+1} J^{n+1} d\widehat{\Omega}
  -\int_{\widehat{\Omega}_e}\textbf{U}^n\phi_j^n J^n d\widehat{\Omega}
  \right]
  +
  \int_{\widehat{\Gamma}_e}\phi_j^{n+1}  (\textbf{F}^{n+1}-\textbf{U}^{n+1}\textbf{V}_g^{n+1})\cdot \textbf{n} J^{n+1} d\widehat{\Gamma}
  \\
  -
  \int_{\widehat{\Omega}_e} (\textbf{F}^{n+1}-\textbf{U}^{n+1}\textbf{V}_g^{n+1}) \cdot\nabla\phi_j^{n+1} J^{n+1} d\widehat{\Omega}
  -
  \int_{\widehat{\Omega}_e}\textbf{U}^{n+1}\frac{d \phi_j^{n+1}}{d t}   J^{n+1} d\widehat{\Omega}
  = 0
\end{aligned}
\end{split}
\end{equation}

and for space-time DG
\begin{equation}
\begin{split}
\begin{aligned}
  \int_{\widehat{\Omega}_e}\textbf{U}^{n+1}\phi_j^{n+1} 
  J^{n+1} d\widehat{\Omega}
  -\int_{\widehat{\Omega}_e}\textbf{U}^n\phi_j^n 
  J^n d\widehat{\Omega}
  +
  \int_t^{t+\Delta t}
  \int_{\widehat{\Gamma}_e}\phi_j  (\textbf{F}-\textbf{U}\textbf{V}_g)\cdot \textbf{n} 
  J d\widehat{\Gamma} dt
  \\
  -
  \int_t^{t+\Delta t}
  \int_{\widehat{\Omega}_e} (\textbf{F}-\textbf{U}\textbf{V}_g) \cdot\nabla\phi_j 
  J d\widehat{\Omega} dt
  -
  \int_t^{t+\Delta t}
  \int_{\widehat{\Omega}_e}\textbf{U}\frac{d \phi_j}{d t}   
  J d\widehat{\Omega} dt
  = 0
\end{aligned}
\end{split}
\end{equation}

Also the two GCL equations corresponding to the above discretizations can be rewritten accordingly as
\begin{equation}
\begin{split}
\begin{aligned}
  \frac{1}{\Delta t}\left[
  \int_{\widehat{\Omega}_e}\phi_j^{n+1} J^{n+1}d\widehat{\Omega}
  -\int_{\widehat{\Omega}_e}\phi_j^n J^nd\widehat{\Omega}
  \right]
  =
  \int_{\widehat{\Gamma}_e}\phi_j^{n+1}  \textbf{V}_g^{n+1}\cdot \textbf{n} J^{n+1}d\widehat{\Gamma} dt
  \\
  -
  \int_{\widehat{\Omega}_e} \textbf{V}_g^{n+1} \cdot\nabla\phi_j^{n+1} J^{n+1}d\widehat{\Omega} dt
  +
  \int_{\widehat{\Omega}_e}\frac{d \phi_j^{n+1}}{d t}   J^{n+1}d\widehat{\Omega} dt
\end{aligned}
\end{split}
\end{equation}
and
\begin{equation}
\begin{split}
\begin{aligned}
  \int_{\widehat{\Omega}_e}\phi_j^{n+1} J^{n+1}d\widehat{\Omega}
  -\int_{\widehat{\Omega}_e}\phi_j^n J^nd\widehat{\Omega}
  =
  \int_t^{t+\Delta t}
  \int_{\widehat{\Gamma}_e}\phi_j  \textbf{V}_g\cdot \textbf{n} Jd\widehat{\Gamma} dt
  \\
  -
  \int_t^{t+\Delta t}
  \int_{\widehat{\Omega}_e} \textbf{V}_g \cdot\nabla\phi_j Jd\widehat{\Omega} dt
  +
  \int_t^{t+\Delta t}
  \int_{\widehat{\Omega}_e}\frac{d \phi_j}{d t}   Jd\widehat{\Omega} dt
\end{aligned}
\end{split}
\end{equation}

And the constraints we want to enforce can be rewritten as well
\begin{equation}
\begin{split}
\left\{
\begin{aligned}
  \int_{\widehat{\Gamma}_e}\phi_j^{n+1} \textbf{V}_g^{n+1}\cdot \textbf{n} J^{n+1}d\widehat{\Gamma}
  &=
  \frac{1}{\Delta t} \int_t^{t+\Delta t}
  \int_{\widehat{\Gamma}_e}\phi_j  \textbf{V}_g\cdot \textbf{n} Jd\widehat{\Gamma} dt
  \\
  \int_{\widehat{\Omega}_e} \textbf{V}_g^{n+1} \cdot\nabla\phi_j^{n+1} J^{n+1}d\widehat{\Omega}
  &=
  \frac{1}{\Delta t} \int_t^{t+\Delta t}
  \int_{\widehat{\Omega}_e} \textbf{V}_g \cdot\nabla\phi_j Jd\widehat{\Omega} dt
  \\
  \int_{\widehat{\Omega}_e}\frac{d \phi_j^{n+1}}{d t}   J^{n+1}d\widehat{\Omega}
  &=
  \frac{1}{\Delta t} \int_t^{t+\Delta t}
  \int_{\widehat{\Omega}_e}\frac{d \phi_j}{d t}   Jd\widehat{\Omega} dt
\end{aligned}
\right.
\end{split}
\end{equation}
We require the following conditions hold at each Gauss quadrature point
\begin{equation}
\begin{split}
\left\{
\begin{aligned}
  \left[
  \phi_j \textbf{V}_g\cdot \textbf{n} J
  \right]^{n+1}
  &=
  \frac{1}{\Delta t} \int_t^{t+\Delta t}
  \phi_j  \textbf{V}_g\cdot \textbf{n} Jdt
  \\
  \left[
   \textbf{V}_g \cdot\nabla\phi_j J
   \right]^{n+1}
  &=
  \frac{1}{\Delta t} \int_t^{t+\Delta t}
   \textbf{V}_g \cdot\nabla\phi_j J dt
  \\
  \left[
  \frac{d \phi_j}{d t} J
   \right]^{n+1}
  &=
  \frac{1}{\Delta t} \int_t^{t+\Delta t}
  \frac{d \phi_j}{d t}   J dt
\end{aligned}
\right.
\end{split}
\end{equation}
For the computation of the total time derivative of the basis 
functions, the finite difference could be used.
After plugging these grid velocity terms into Eq. \ref{eq:GCL:BDF1:U}, 
the right-hand side could be evaluated at time level $n+1$.

For the higher-order ESDIRK temporal scheme (Explicit first stage, 
Single Diagonal coefficient, diagonally Implicit Runge-Kutta) 
\cite{wang2007implicit,xia2015third,wang2017nka}, we have similarly
\begin{equation}
\begin{split}
\begin{aligned}
  \frac{1}{\Delta t}\left[
  \int_{\widehat{\Omega}_e}\textbf{U}^{s}\phi_j^{s} J^{s} d\widehat{\Omega}
  -\int_{\widehat{\Omega}_e}\textbf{U}^n\phi_j^n J^n d\widehat{\Omega}
  \right]
  +
  \sum_{r=1}^{S}\alpha_{sr}
  \int_{\widehat{\Gamma}_e} 
  \left[\phi_j  (\textbf{F}-\textbf{U}\textbf{V}_g)\cdot \textbf{n} J \right]^r
   d\widehat{\Gamma}
  \\
  -
  \sum_{r=1}^{S}\alpha_{sr}
  \int_{\widehat{\Omega}_e} 
  \left[ (\textbf{F} -\textbf{U}\textbf{V}_g) \cdot\nabla\phi_j J \right]^r
  d\widehat{\Omega}
  -
  \sum_{r=1}^{S}\alpha_{sr}
  \int_{\widehat{\Omega}_e}
  \left[ \textbf{U}\frac{d \phi_j}{d t} J \right]^r
  d\widehat{\Omega}
  = 0
  \hspace{0.2in}
  s = 1,2,...,S
\end{aligned}
\end{split}
\end{equation}
where each stage corresponds to time $t^s = t^n + c_s\Delta t$.
Similarly, by comparing each stage of the ESDIRK scheme with the space-time DG formulation,
the following conditions could be obtained, such that each stage will satisfy the GCL.
\begin{equation}
\begin{split}
\left\{
\begin{aligned}
  \sum_{r=1}^{S}\alpha_{sr}
  \left[
  \phi_j \textbf{V}_g\cdot \textbf{n} J
  \right]^r
  &=
  \frac{1}{\Delta t} \int_t^{t+c_s\Delta t}
  \phi_j  \textbf{V}_g\cdot \textbf{n} Jdt
  \\
  \sum_{r=1}^{S}\alpha_{sr}
  \left[
   \textbf{V}_g \cdot\nabla\phi_j J
   \right]^r
  &=
  \frac{1}{\Delta t} \int_t^{t+c_s\Delta t}
   \textbf{V}_g \cdot\nabla\phi_j J dt
  \\
  \sum_{r=1}^{S}\alpha_{sr}
  \left[
  \frac{d \phi_j}{d t} J
   \right]^r
  &=
  \frac{1}{\Delta t} \int_t^{t+c_s\Delta t}
  \frac{d \phi_j}{d t}   J dt
\end{aligned}
\right.
\end{split}
\hspace{0.4in} s = 1,2,...,S
\end{equation}
Note that these conditions are required for each Gauss quadrature point. 
For the ESDIRK scheme, this system of equations can be easily solved by forward substitution. 
We choose the third order ESDIRK3 scheme for this work.

\section{Mesh Motion}
\label{sec:meshMotion}

In both the curved mesh generation and the pure mesh movement, the motion of the 
boundary nodes are required to propagate to the interior ones properly, 
in order to avoid invalid elements near the boundary and to maintain good mesh quality. 
That is, given the displacements of the boundary nodes, it is desired to obtain 
the displacements for all the interior nodes.
With the motion of all the nodes in the mesh determined, one can then move the mesh coordinates
accordingly to accommodate the physics solver, e.g., the ALE formulation for the fluid flow.

There are several types of methods in the literature dealing with the movement of the interior nodes. 
Model based on physics (solid mechanics) is a popular one. 
Examples are linear elasticity \cite{mm:curve:hartmann}, non-linear elasticity \cite{mm:curve:persson},
linear spring analogy \cite{mm:batina} and torsion spring analogy \cite{mm:torsionsprings}. 
This type of method requires the solution of a system of equations, 
elliptic (Poisson-type) partial differential equations (PDE) in the case of the linear elasticity model,
thus is relatively computationally expensive. 
The second type is the interpolation method or algebraic method, 
including radial basis function (RBF) interpolation \cite{mm:rbf}, 
explicit interpolation \cite{mm:explicit} and Delaunay graph mapping \cite{mm:dgm}.
Kashi and Luo \cite{kashi2016aiaa,kashiThesis} compared the performance between RBF interpolation and some of the other methods. 
It is found that the RBF method, based on the interpolation technique 
which is relatively fast, can give good and robust results in general, especially for large or rotational deformations.

In the following moving airfoil test cases, the displacement or grid velocity of the wall boundary are provided in advance. 
The RBF interpolation method is then used to compute the deformation of
the interior nodes, such that the nodes on the airfoil are in a rigid motion, 
and those far away from the wall are kept static.
We note that if a curved mesh is to be used, then the high order nodes 
(e.g., midpoint of an edge in 2D) will also come into play.

Here we briefly recall the methodology of the RBF method. 
Interested readers are encouraged to refer to  \cite{kashi2016aiaa,kashiThesis} and references therein for more information.
The RBF method states that, given $N_b$ boundary points in the mesh, 
we require the displacement at any point $\mathbf{x}$ (both boundary nodes and the interior nodes) in the mesh satisfy
\begin{equation}
  \label{eq:rbf:1}
  \mathbf{s}(\mathbf{x}) = \sum_{j=1}^{N_b} \mathbf{a}_j\phi(\lVert \mathbf{x}-\mathbf{x}_{bj}\rVert)
\end{equation}
where $\phi(r)$ is the basis function, 
$\mathbf{s}$ represents the displacement vector field with 2 components x- and y- in 2D ( x-, y- and z- in 3D), 
and $\mathbf{a}_j$ are the coefficients to be determined. 
Note that the RBF method will be applied to a scalar field, i.e., to each component individually.
In this work, we choose the modified Wendland's C2 function \cite{mm:rbf:errorwendland}
as the basis function
\begin{equation}
 \phi(r) = 
 \begin{cases}
   \left(1-\frac{r}{r_s}\right)^4\left(4\frac{r}{r_s} + 1\right) & r \leq r_s \\
   0 & r > r_s
 \end{cases}
\end{equation}
where $r_s$ is called the support radius, indicating how far the boundary deformation will propagate to the interior.
Given the displacement of the boundary nodes, by substituting these 
known coordinates and displacements into Eq. \ref{eq:rbf:1}, we can 
construct a linear system of equations for the coefficients $\textbf{a}_j$ in each dimension in space
\begin{equation}
  \textbf{s}(\textbf{x}_{bi}) = \sum_{j=1}^{N_b} \textbf{a}_j\phi(\lVert \textbf{x}_{bi}-\textbf{x}_{bj}\rVert)
\end{equation}
or rewritten as
\begin{equation}
  \textbf{A}\textbf{a}=\textbf{s}
\end{equation}
When using Wendland's C2 function, the resulting symmetric system is 
claimed to be positive definite \cite {mm:rbf:errorwendland}, thus can 
be easily solved by conjugate gradient (CG) or Choleskey decomposition methods, or even direct methods. 
After solving the linear system, we can evaluate the displacement for any interior nodes using Eq. \ref{eq:rbf:1}.

  \section{ Numerical Examples }
  \label{sec:numerical-examples}
In all the following test cases, the 3rd-order accurate temporal 
discretization ESDIRK3 scheme is used, and the curved elements ($Q2$) 
are chosen for all the grids, in order to test the 
applicability of the rDG-ALE method.

\subsection{Uniform Flow}
\begin{figure}[H]
  \centering
  \subfloat[t=0]{  
    \includegraphics[trim= 1.8cm 1.5cm 4.2cm 2cm,clip,height = 2in]
    {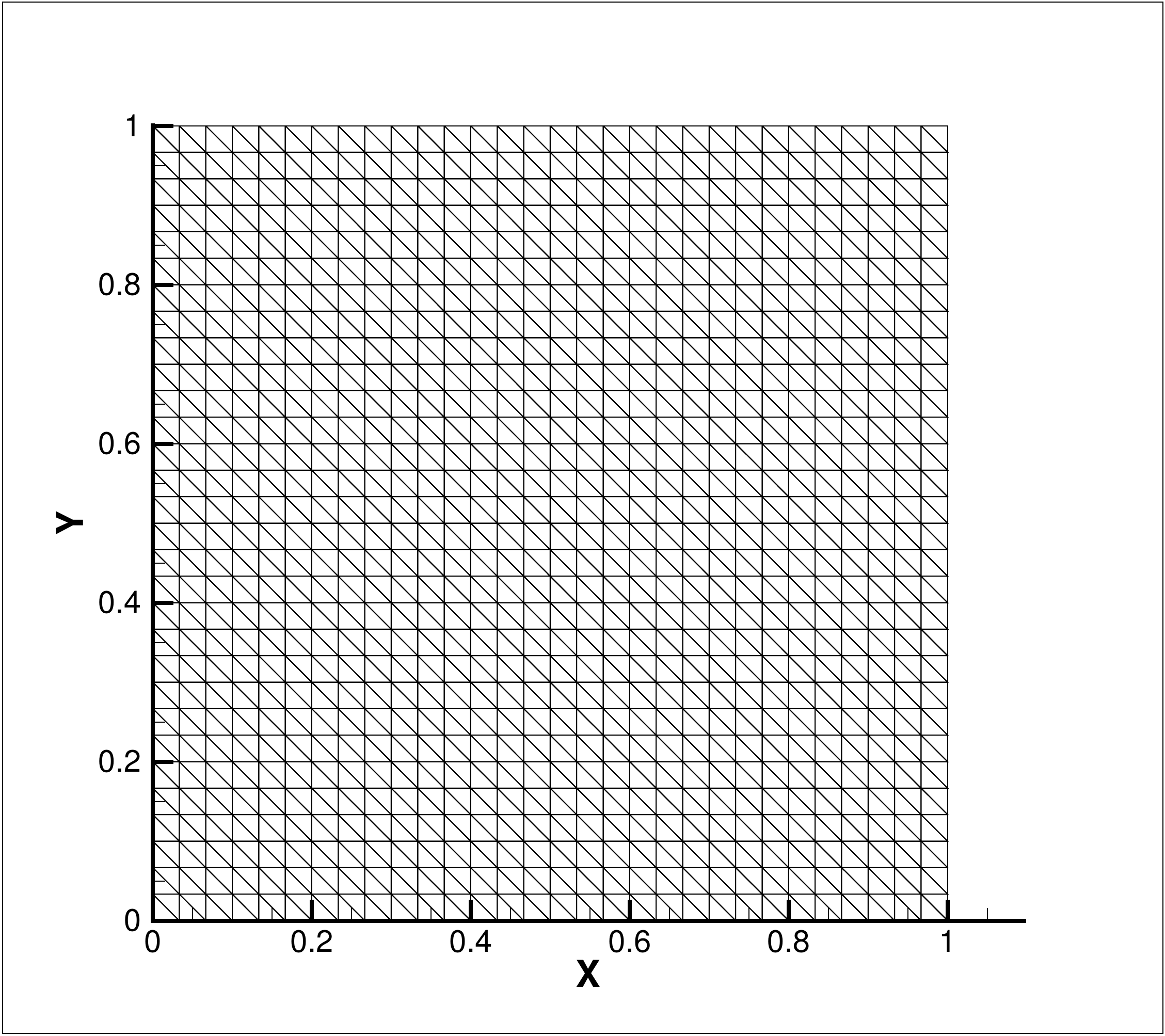} }
  \subfloat[t=0.3T]{  
    \includegraphics[trim= 2cm 1.5cm 4.2cm 2cm,clip,height = 2in]
    {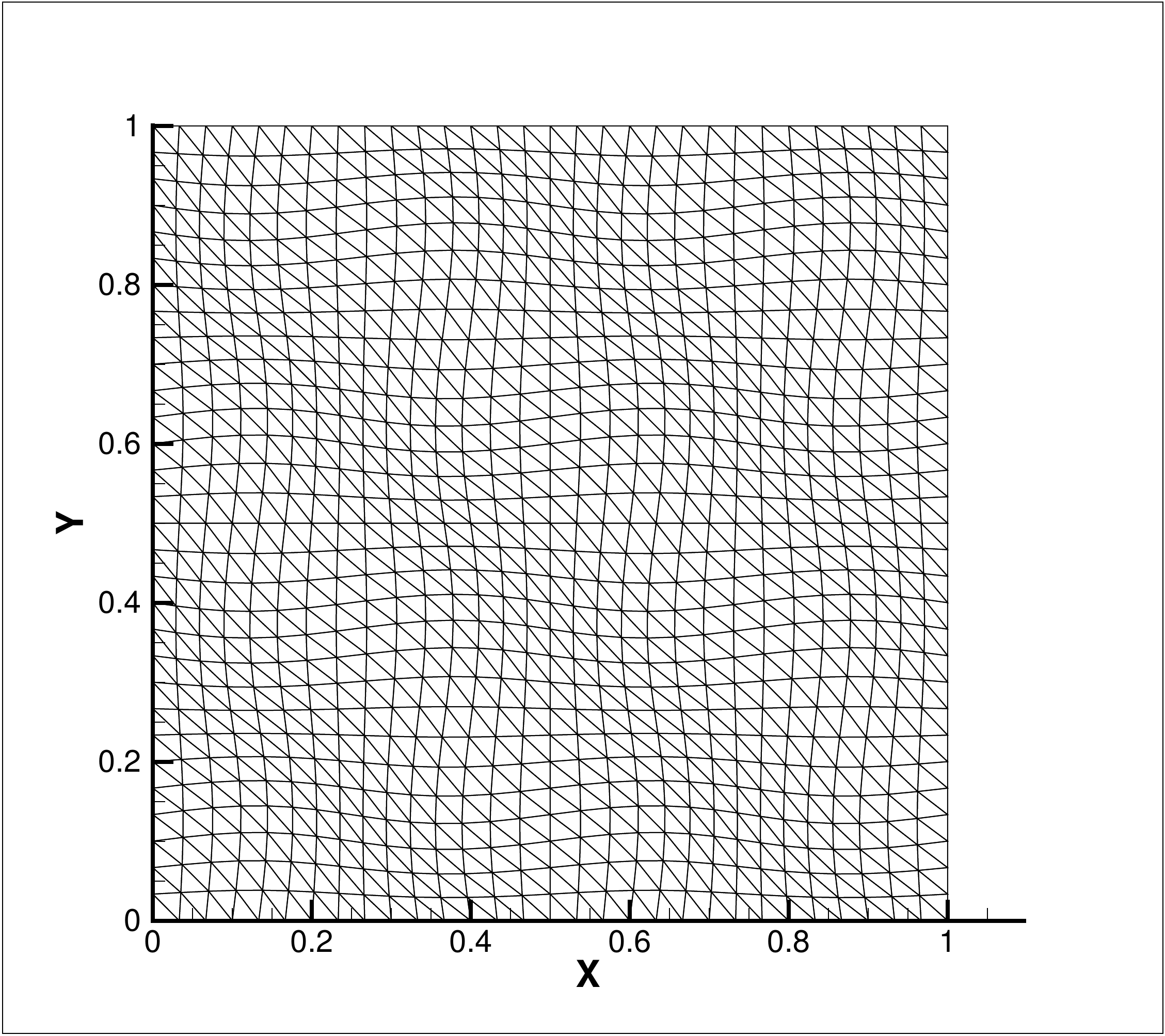} }
  \\
  \subfloat[t=0.7T]{  
    \includegraphics[trim= 1.8cm 1.5cm 4.2cm 2cm,clip,height = 2in]
    {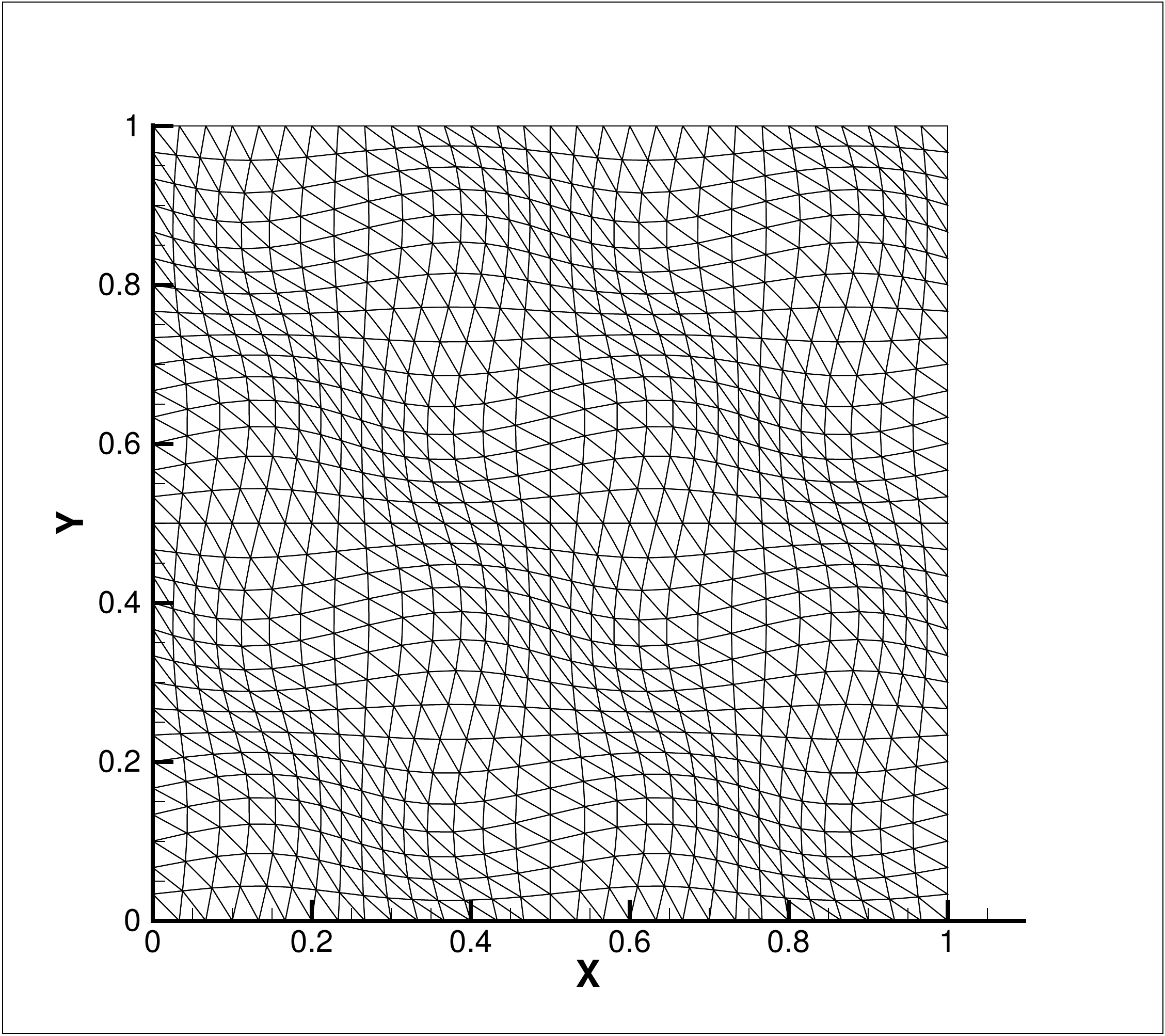} }
  \subfloat[t=1.0T]{  
    \includegraphics[trim= 2cm 1.5cm 4.2cm 2cm,clip,height = 2in]
    {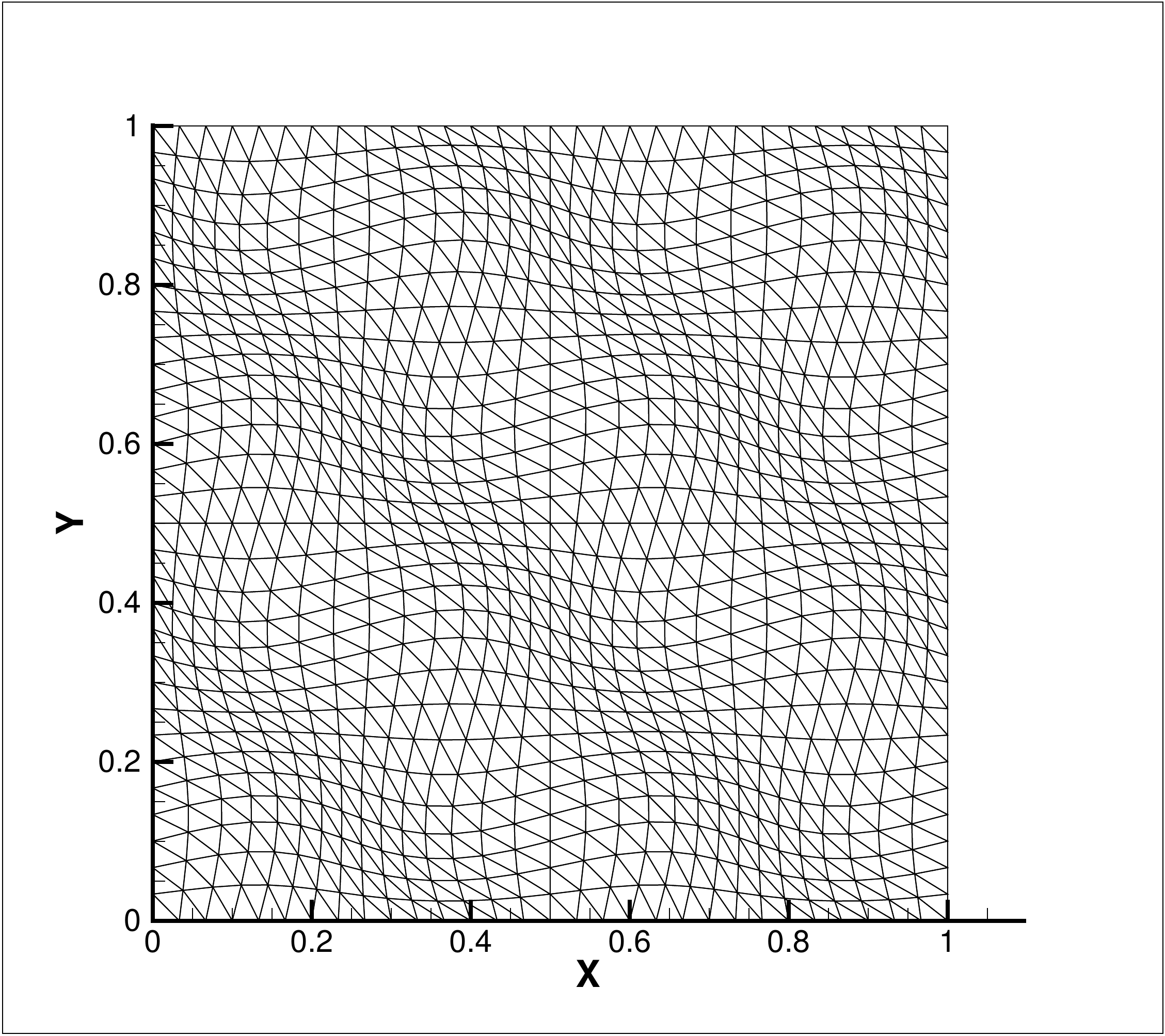} }
  \caption{Mesh deformation for the uniform flow test case.}
  \label{fig:uniformFlow:mesh}
\end{figure}

The first test case is the uniform flow on deforming grids. 
The initial mesh is uniform on a square domain defined in $0\leqslant x,y \leqslant 1.0$, and the mesh motion is prescribed by
$$\textbf{x}(t) = \textbf{x}_0 + \textbf{dx}(t) $$
where
$$ dx(t) = \frac{A_xL_x}{T}\text{sin}(\frac{n_t\pi t}{T})\text{sin}(\frac{n_x\pi x}{L_x})\text{sin}(\frac{n_y\pi y}{L_y})$$
$$ dy(t) = \frac{A_yL_y}{T}\text{sin}(\frac{n_t\pi t}{T})\text{sin}(\frac{n_x\pi x}{L_x})\text{sin}(\frac{n_y\pi y}{L_y})$$
where $L_x = L_y = 1.0$ and $T=1.0$ are the reference length and time,  $A_x=A_y=0.025$ is the deformation amplitude, $n_x=n_y=4$ and $n_t=0.5$ are for the deformation frequency in space and time.
The mesh deformation process is illustrated in Fig. \ref{fig:uniformFlow:mesh}. 
The $L_2$ norm of the numerical error is measured and observed to be around $10^{-12}$, thus verifying the Geometric Conservation Law (GCL).

\subsection{Convection of an Isentropic Vortex}

The convection of a 2D inviscid isentropic vortex is considered in this test case. 
All the three spatial discretization methods, DG(P1), rDG(P1P2) and DG(P2) are employed in the computations. 
The square domain $0\leqslant x,y \leqslant 1.0$ is considered for 
this test case as well, with the same mesh deformation configuration as in the uniform flow case.

The mean flow is set as $(\rho_{\infty},u_{\infty},v_{\infty},p_{\infty},T_{\infty})=(1.0,0.5,0.0,1.0,1.0)$. At $t_0=0$, the flow is perturbed by an isentropic vortex centered at $(x_0,y_0) = (0.25,0.25) $ with
\begin{equation}
\begin{split}
  \delta T &= -\frac{\alpha^2(\gamma-1)}{16\phi\gamma\pi^2}e^{2\phi(1-r^2)},\\
  \delta u &= -\frac{\alpha}{2\pi}(y-y_0)e^{\phi(1-r^2)},\\
  \delta v &= \frac{\alpha}{2\pi}(x-x_0)e^{\phi(1-r^2)},
\end{split}
\end{equation}
where $r=\sqrt{(x-x_0)^2+(y-y_0)^2}$ and $\phi=1.0, \alpha=4.0, \gamma=1.4$.

The other conservative variables can be determined by the follows:
\begin{equation}
\begin{split}
  \rho &= (T_{\infty}+\delta T)^{1/(\gamma-1)},\\
  u    &= u_{\infty} + \delta u,\\
  v    &= v_{\infty} + \delta v,\\
  p    &= \rho^{\gamma}.
\end{split}
\end{equation}

\begin{figure}[H]
  \centering
  \subfloat{
    \includegraphics[trim= 2cm 1.5cm 0cm 2cm,clip,height = 2in]
    {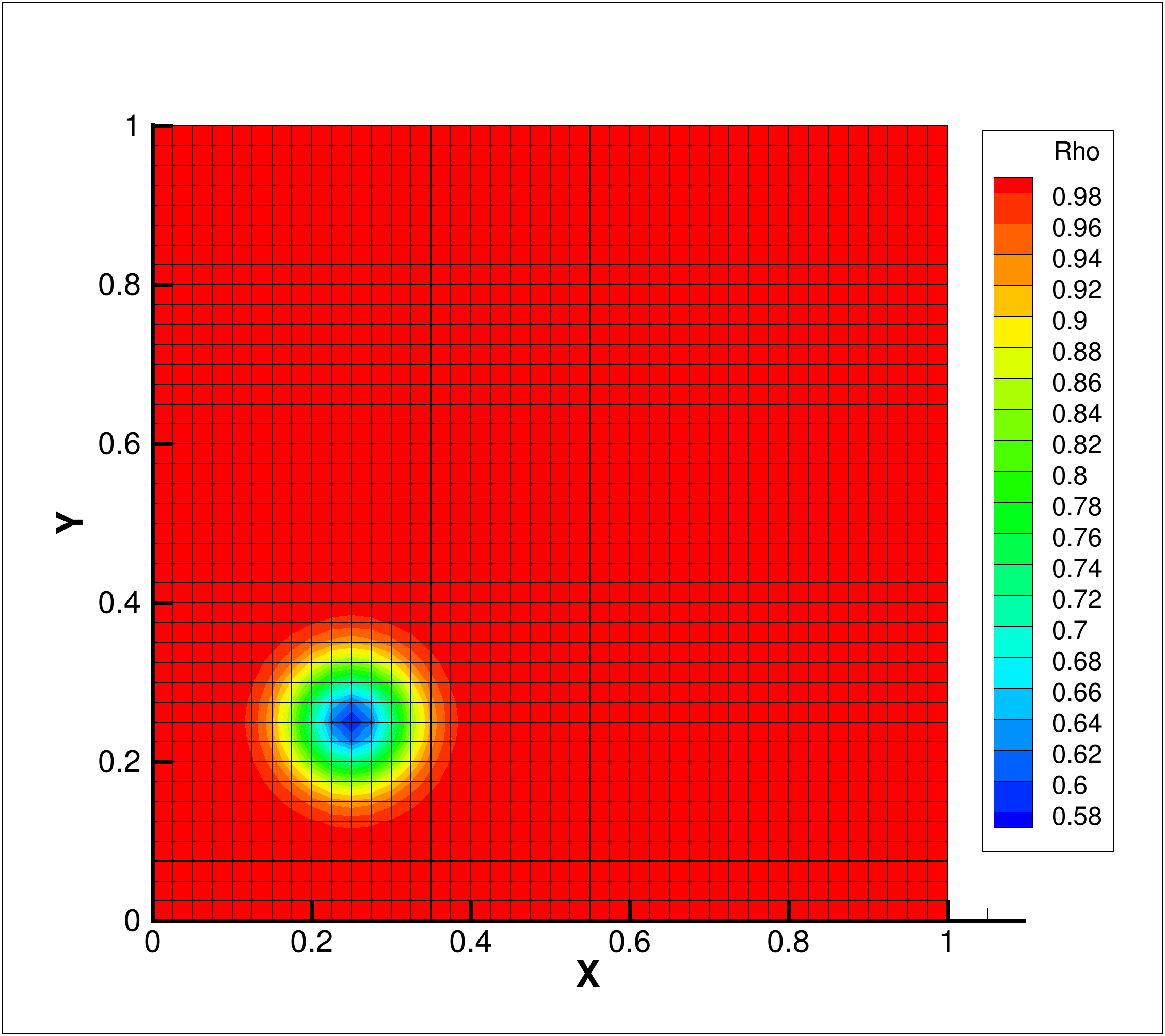}}  
  \caption{Initial mesh and density distribution for the isentropic vortex problem.}
  \label{fig:isentropicVortex:initialMesh}
\end{figure}

\begin{figure}[H]
  \centering
  \subfloat[final-DG(P1)]{
    \includegraphics[trim= 1.8cm 1.5cm 1cm 2cm,clip,height = 2in]
    {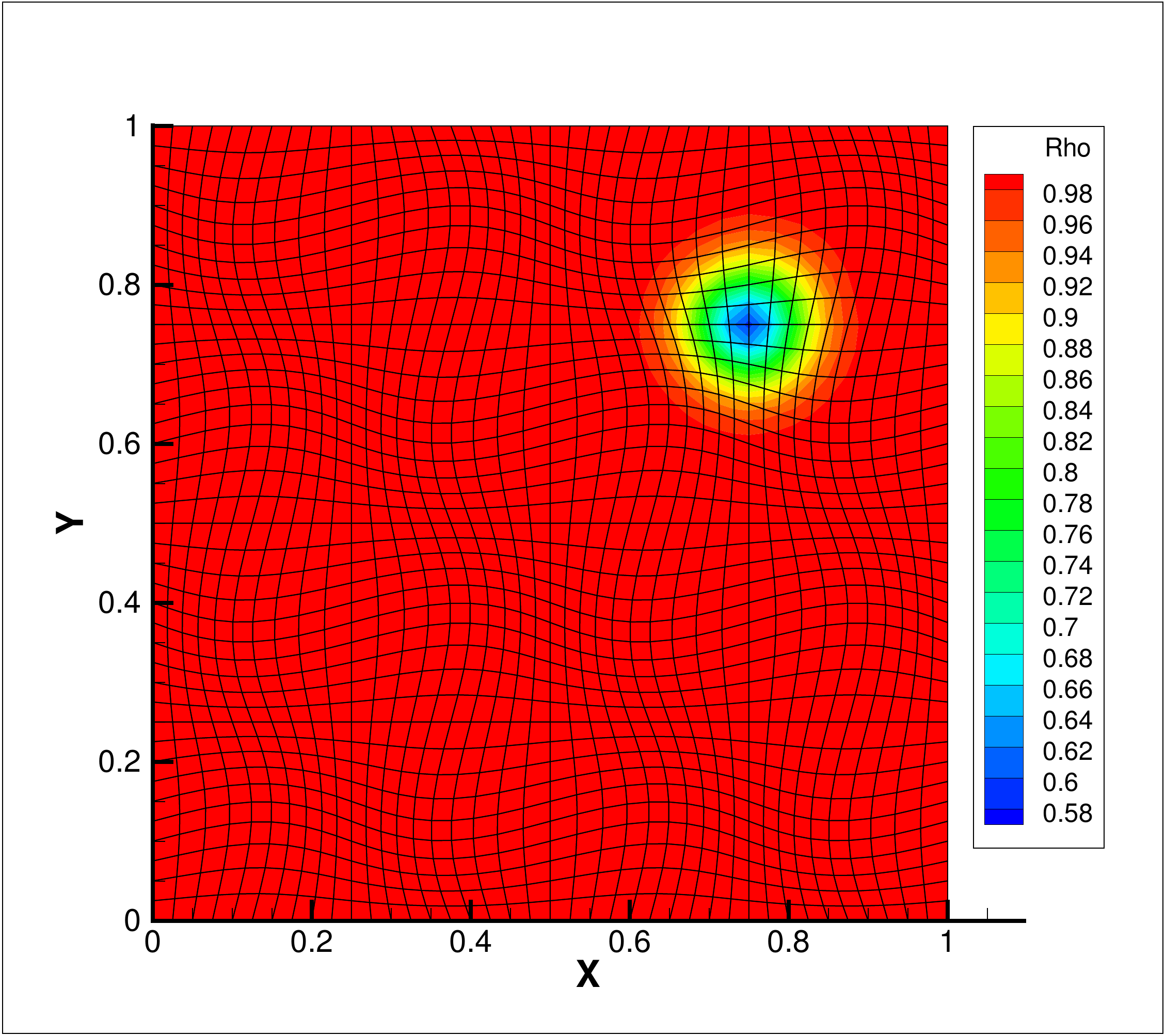}}
  \subfloat[final-rDG(P1P2)]{
    \includegraphics[trim= 1.8cm 1.5cm 1cm 2cm,clip,height = 2in]
    {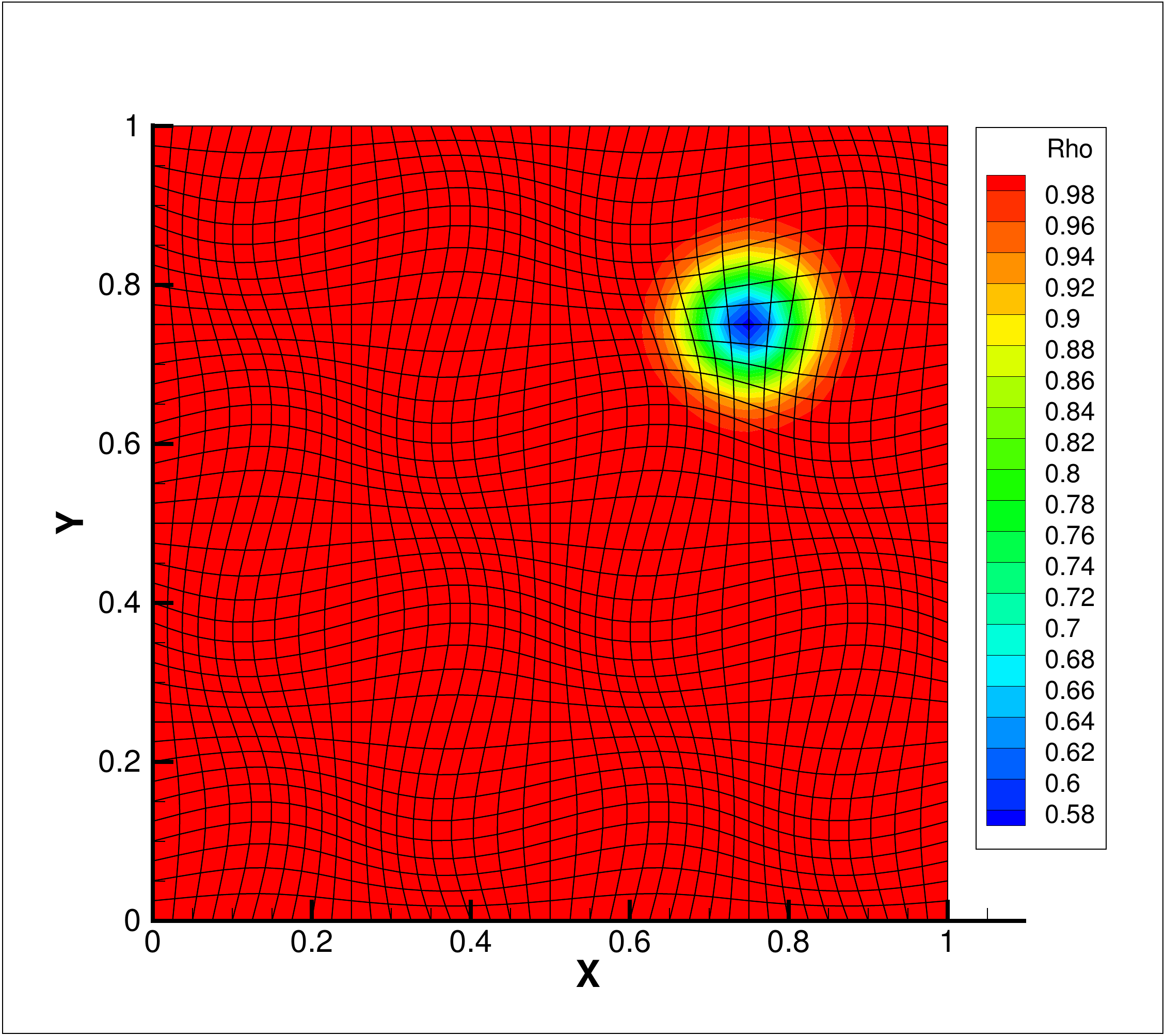}}
  \caption{Computed density for DG(P1) and rDG(P1P2) discretizations of the isentropic vortex problem.}
  \label{fig:isentropicVortex:densityFinal}
\end{figure}

\begin{figure}[H]
  \centering
  \includegraphics[height =2.5in]
    {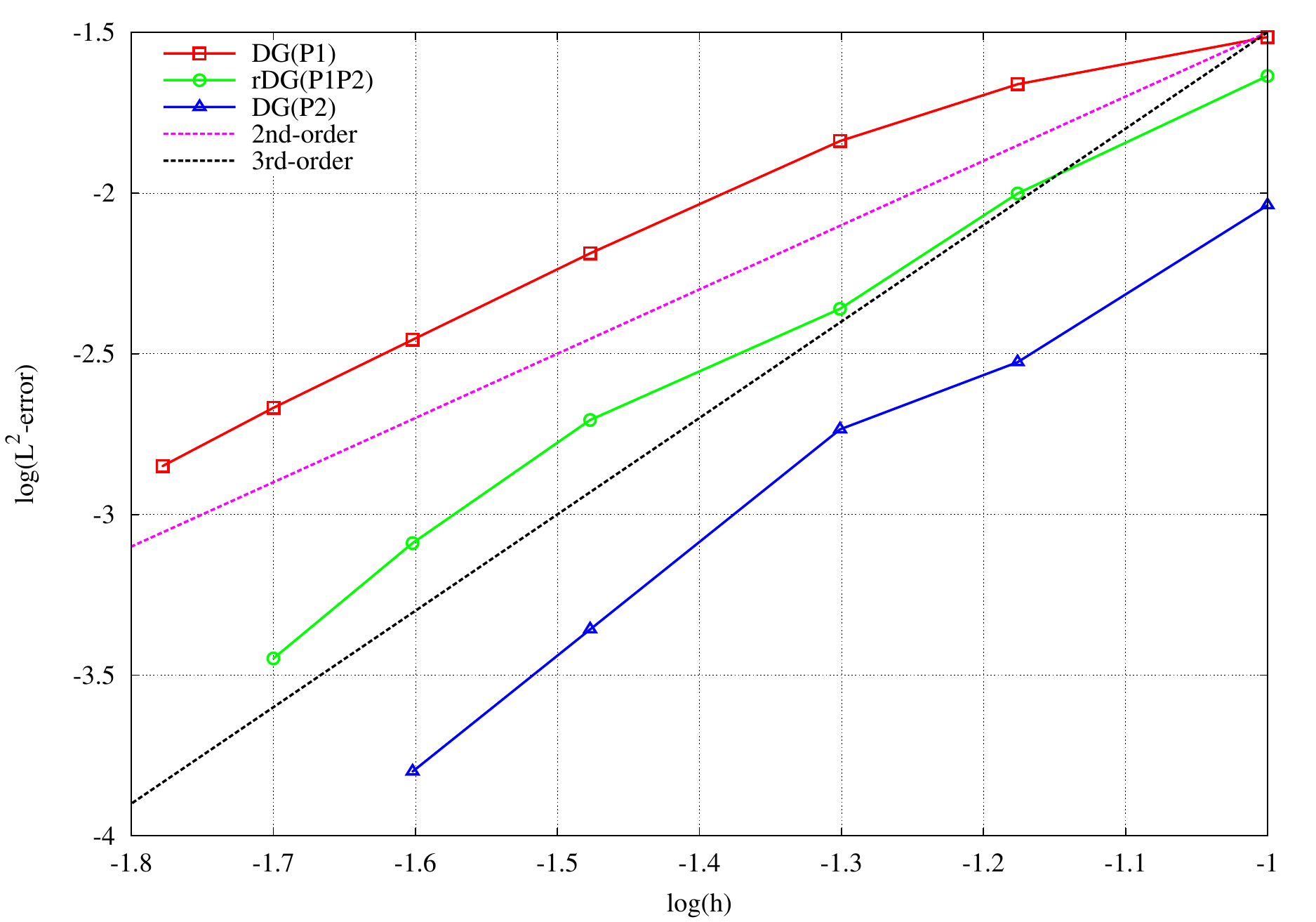}
  \caption{Spatial convergence rates for DG(P1), rDG(P1P2) and DG(P2) discretizations of the isentropic vortex problem.}
  \label{fig:isentropicVortex:l2norm-spatial}
\end{figure}

\begin{figure}[H]
  \centering
  \includegraphics[height =2.5in]
    {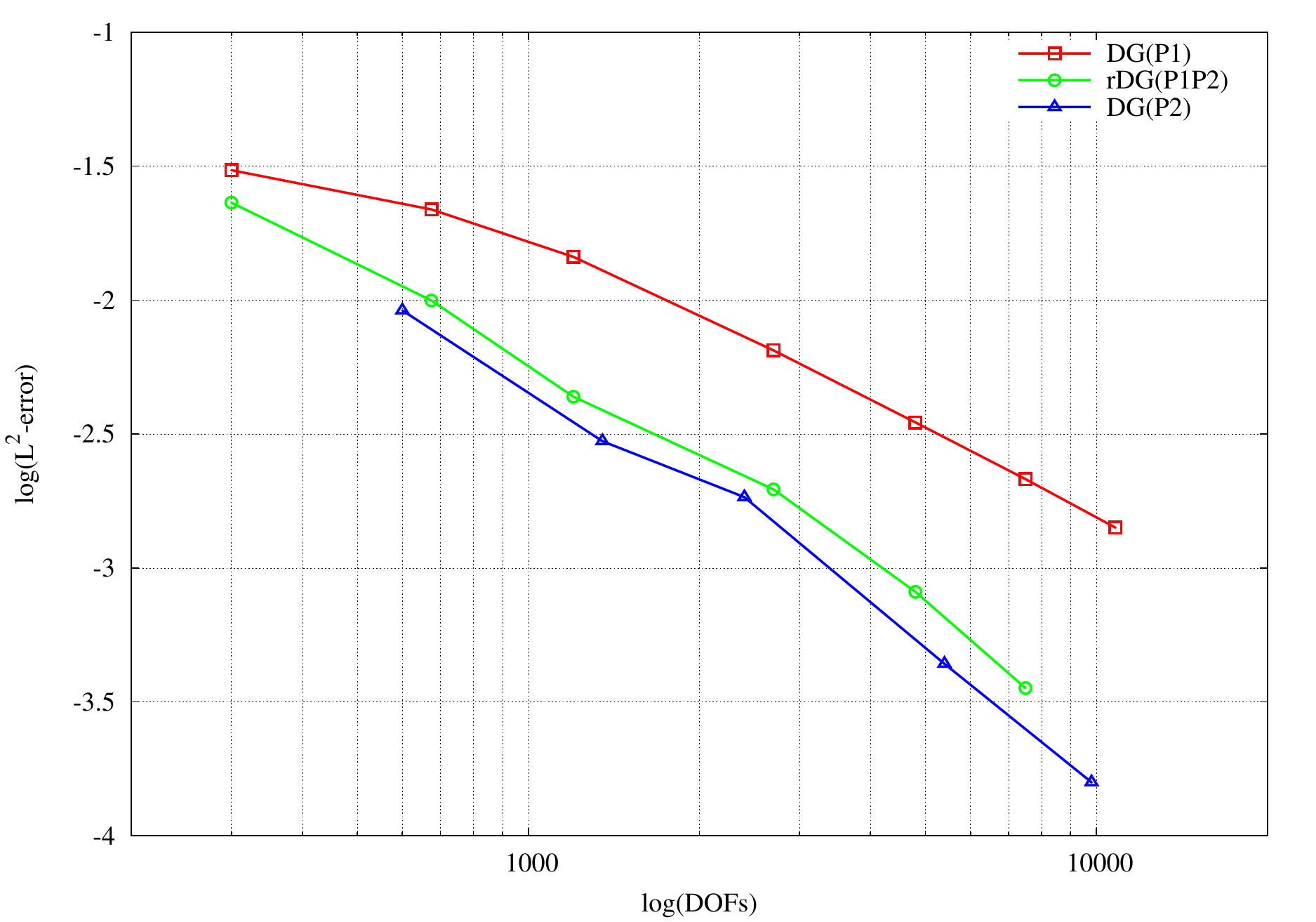}
  \caption{Degrees of freedom versus spatial error for DG(P1), rDG(P1P2) and DG(P2) discretizations of the isentropic vortex problem.}
  \label{fig:isentropicVortex:ndof-spatial}
\end{figure}

The initial mesh and density distribution are shown in Fig. \ref{fig:isentropicVortex:initialMesh}, and the computed solution are shown in Fig. \ref{fig:isentropicVortex:densityFinal}, for DG(P1) and rDG(P1P2), respectively. 
A mesh refinement study is performed, in order to evaluate the convergence rate for different spatial discretization schemes. 
The computed $L_2$ norm for density are shown in Fig. \ref{fig:isentropicVortex:l2norm-spatial}. 
We can see the desired orders of accuracy are achieved, and by comparing rDG(P1P2) with DG(P1), 
we can see rDG(P1P2) not only has higher convergence rate, but also 
the absolute error for rDG(P1P2) is also smaller than the counterpart of DG(P1). 
Fig. \ref{fig:isentropicVortex:ndof-spatial} presents the spatial 
error versus the number of degrees of freedoms (dofs) for DG(P1) and rDG(P1P2). 
One can see that for a given number of dofs, the numerical error of 
rDG(P1P2) is smaller than that of DG(P1), which indicates that rDG(P1P2) is computationally more efficient.

\begin{figure}[H]
  \centering
  \includegraphics[height =2.5in]
    {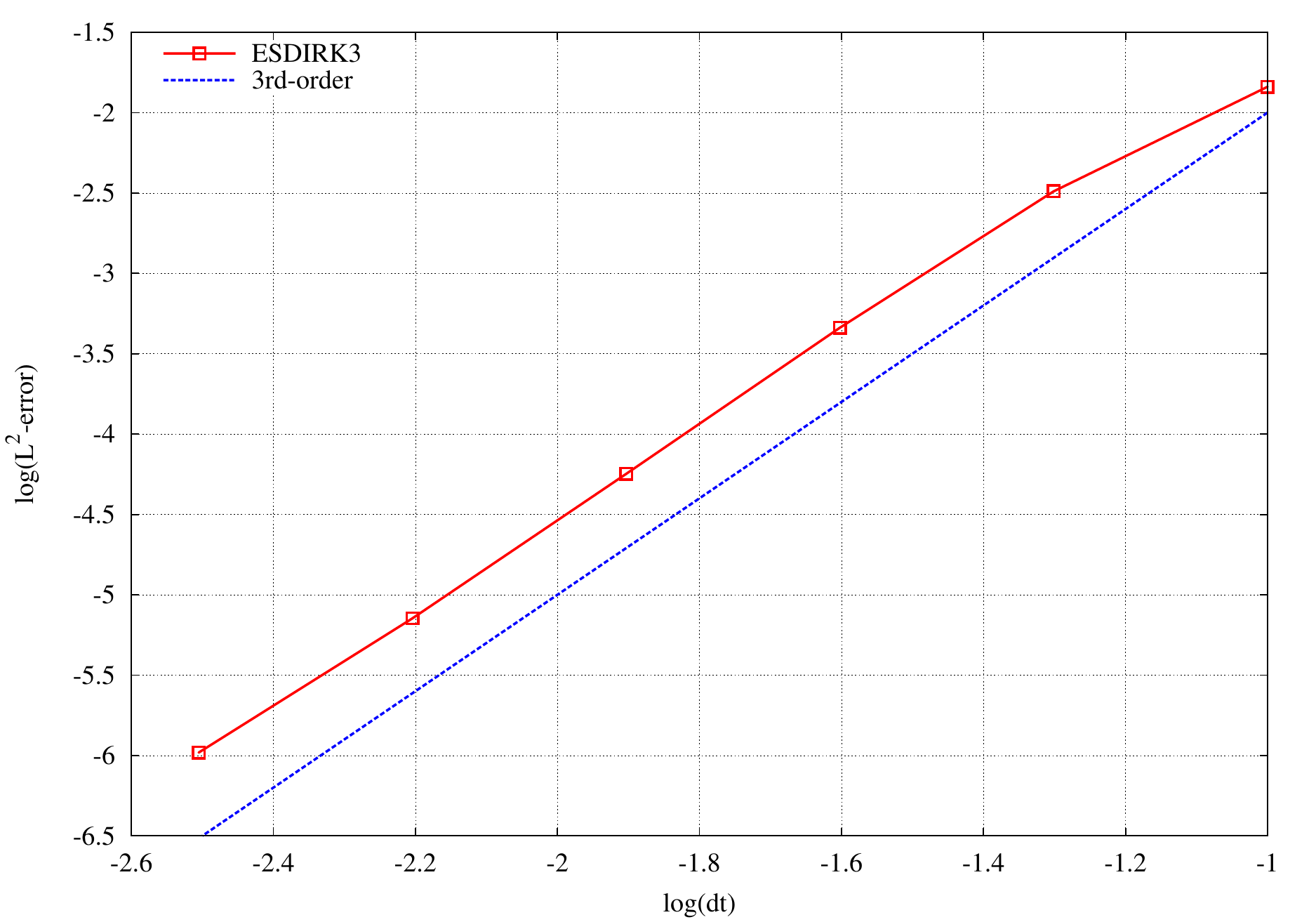}
  \caption{Temporal convergence rate for ESDIRK3 scheme on the isentropic vortex problem.}
  \label{fig:isentropicVortex:l2norm-temporal}
\end{figure}

To demonstrate the temporal order of accuracy of the ESDIRK3 scheme, 
we perform a time-step refinement study, and compute the numerical errors.
The results are shown in Fig. \ref{fig:isentropicVortex:l2norm-temporal}.
We can see that the designed 3rd-order of convergence has been achieved.

\subsection{Oscillatory NACA0012 Airfoil}

\begin{figure}[H]
  \centering
  \includegraphics[trim= 2cm 1.5cm 2.2cm 2cm,clip,height =2in]
    {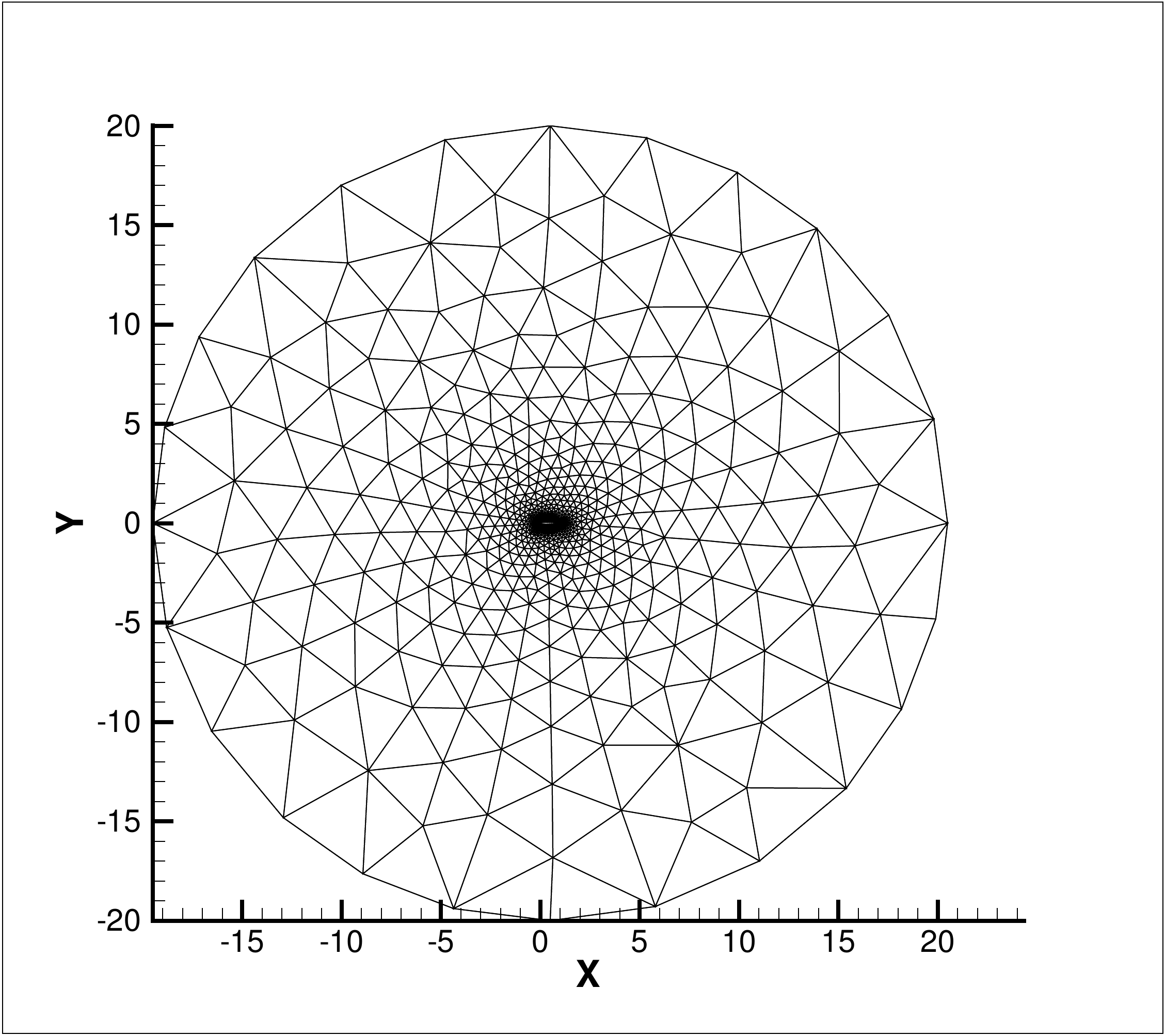}
  \caption{Initial mesh for the pitching NACA0012 case.}
  \label{fig:eulerNaca0012:meshInitialGlobal}
\end{figure}

\begin{figure}[H]
  \centering
  \subfloat[$\alpha = 0.0$]{  
    \includegraphics[trim= 2cm 1.5cm 2.2cm 2cm,clip,height = 2in]
    {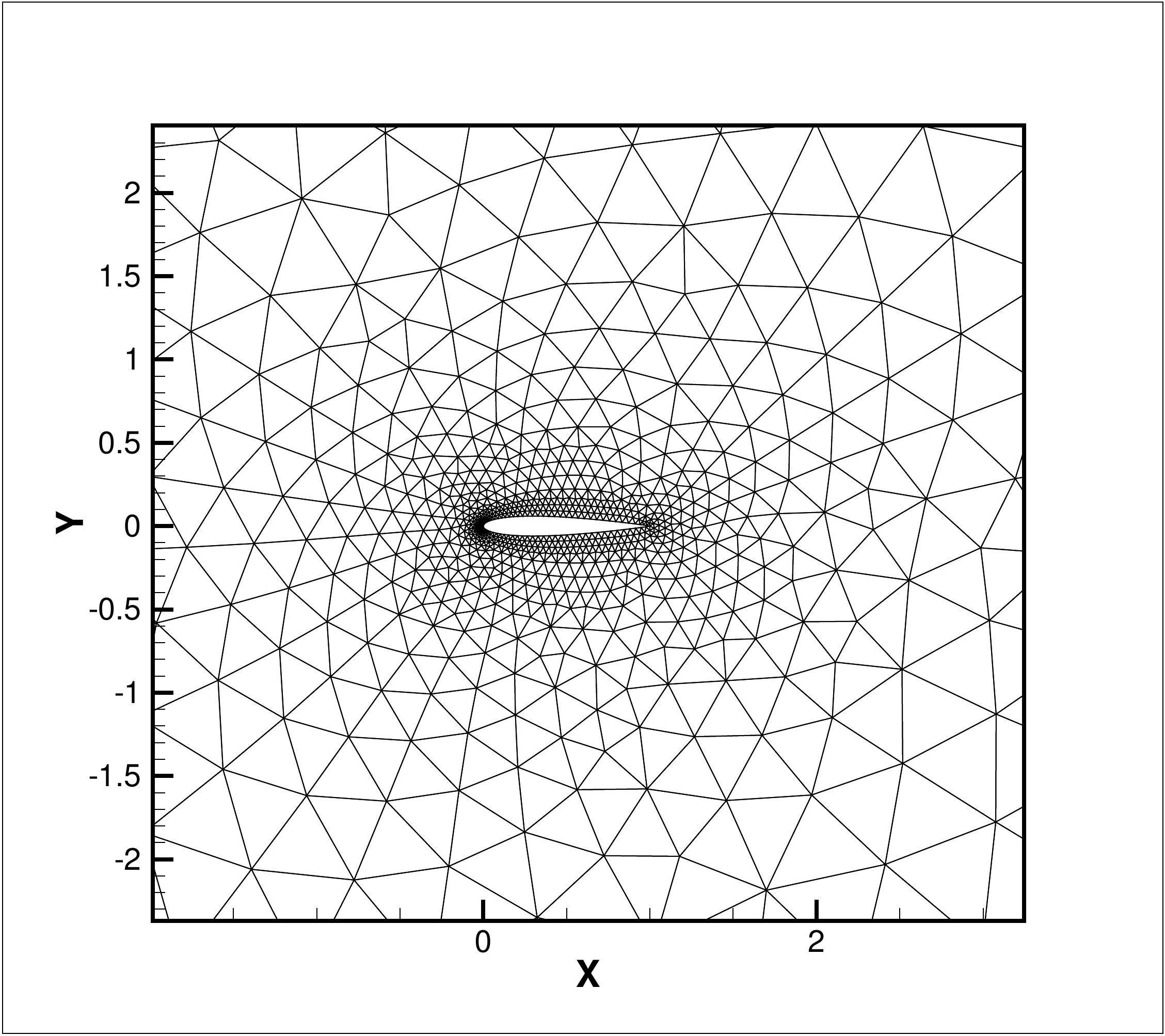} }
  \subfloat[$\alpha = \alpha_m + \alpha_0$]{  
    \includegraphics[trim= 2cm 1.5cm 2.2cm 2cm,clip,height = 2in]
    {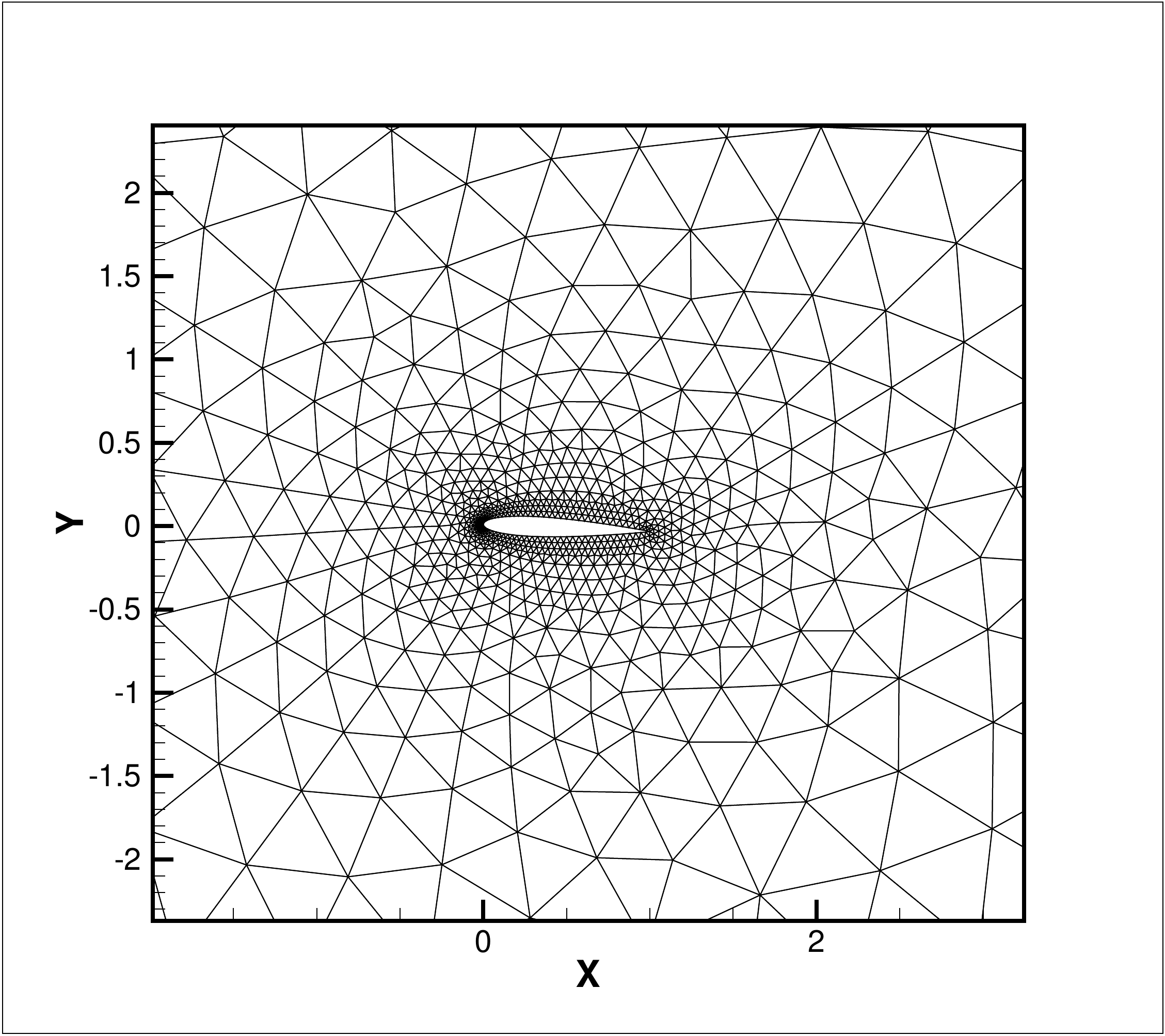} }
  \caption{Zoomed-in mesh near the airfoil. left:initial mesh; right: mesh at maximum angle of attack}
  \label{fig:eulerNaca0012:meshZoomIn}
\end{figure}

Pitching and heaving are typical motions for an airfoil. 
In this test case, we consider the flow over a NACA0012 airfoil harmonically 
pitching about the quarter chord length, 
in order to show the capability of the rDG-ALE method.  
The Mach number of the freestream condition is $\textit{Ma} =0.755$, Reynolds number $\textit{Re}=5.5\times10^6$ 
and the specific heat ratio $\gamma = 1.4$. 
The mesh motion is dictated by the time-dependent angle of attack (AoA) of the airfoil: 
$\alpha(t) = \alpha_m + \alpha_0\text{sin}(\omega t)$, where $\alpha_m = 0.016$ degree 
is the mean incidence, $\alpha_0 = 2.51$ degree the pitching amplitude, 
and the reduced frequency $\omega c/(2U_\infty) = 0.0814$, 
with $c$ the chord length and $U_\infty$ the magnitude of the freestream velocity. 
The computational domain is a circle region with radius $R=20$, and the leading edge of the 
airfoil is at the origin $(x_0,y_0)=(0,0)$, thus pitching axis at $(x_p,y_p)=(0.25,0)$.
The initial mesh and zoomed-in grids near the airfoil at two different angles of attack 
are shown in Fig. \ref{fig:eulerNaca0012:meshInitialGlobal} 
and Fig. \ref{fig:eulerNaca0012:meshZoomIn}, respectively. 
Given the above harmonic motion for the points at wall boundary and the static points at the
far-filed boundary, the movement of the interior nodes are interpolated from the RBF method.
With the airfoil motion and the flow field being periodic, the unsteady solutions after a
reasonably long time of computation should be used for analysis.

As in the experiments by Landon \cite{landonNaca0012}, the normal force coefficients $C_n$ and pitching moment coefficients $C_m$ are defined as\\
\[
C_n = \int_0^1 (C_p^L - C_p^U)d(x/c)
\]

\[
C_m = \int_0^1 (C_p^L - C_p^U)(0.25-(x/c))d(x/c)
\]

where
\[
C_p = \frac{p-p_\infty}{\frac{1}{2}\rho_\infty U_\infty^2}
\]
is the pressure coefficient, and the subscript $\infty$ denotes the freestream quantity, 
while superscript $L$ and $U$ denote the lower and upper surfaces respectively.

\begin{figure}[H]
  \centering
  \subfloat[Normal force coefficient]{  
    \includegraphics[trim= 0cm 0cm 0.5cm 0cm,clip,height = 2.5in]
    {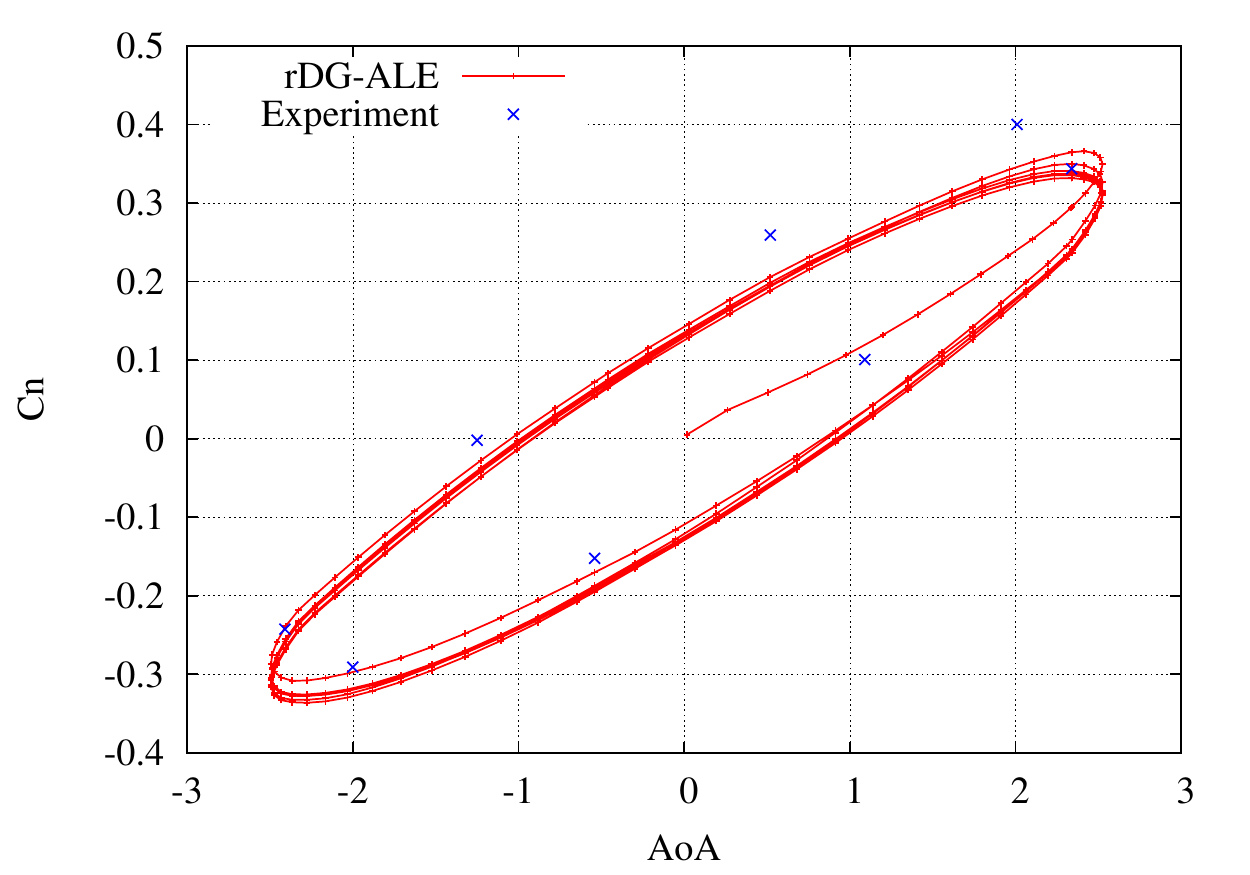} }
  \\
  \subfloat[Pitching moment coefficient]{  
    \includegraphics[trim= 0cm 0cm 0.5cm 0cm,clip,height = 2.5in]
    {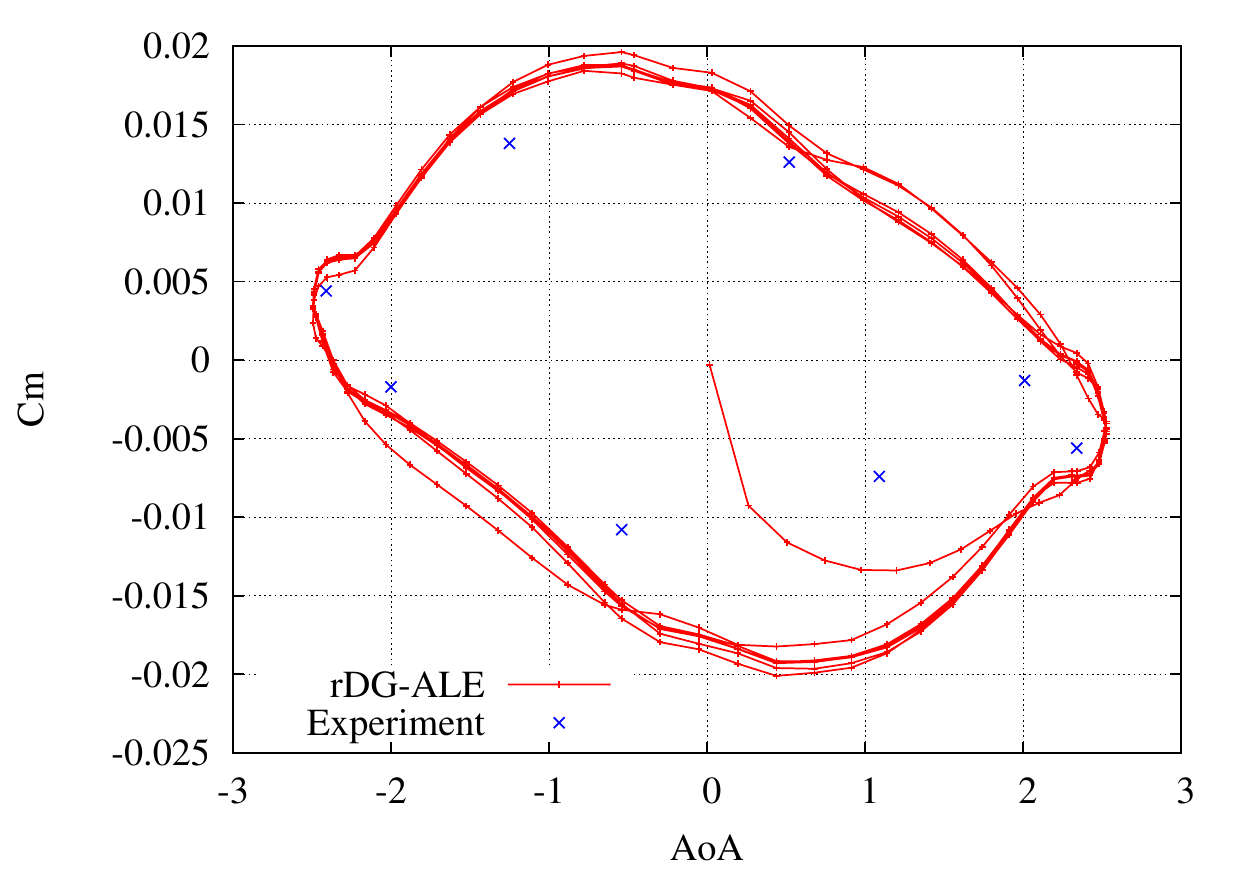} }
  \caption{Normal force coefficients (a) and pitching moment coefficients 
  (b) versus angles of attack (in degree) of the pitching NACA0012 case.}
  \label{fig:eulerNaca0012:clcm-angle}
\end{figure}

\begin{figure}[H]
  \centering
  \subfloat[Normal force coefficient]{  
    \includegraphics[trim= 0cm 0cm 0.5cm 0cm,clip,height = 2.5in]
    {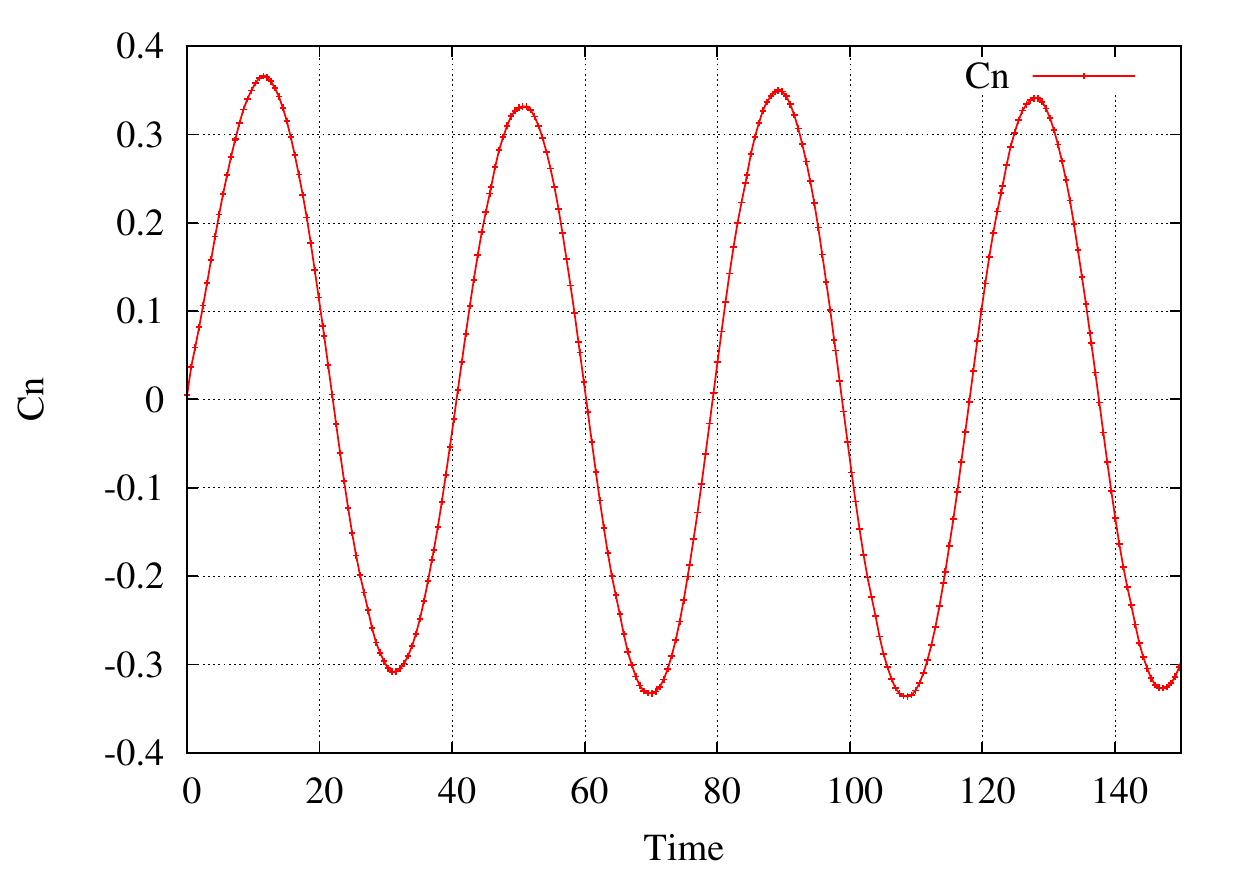} }
  \\
  \subfloat[Pitching moment coefficient]{  
    \includegraphics[trim= 0cm 0cm 0.5cm 0cm,clip,height = 2.5in]
    {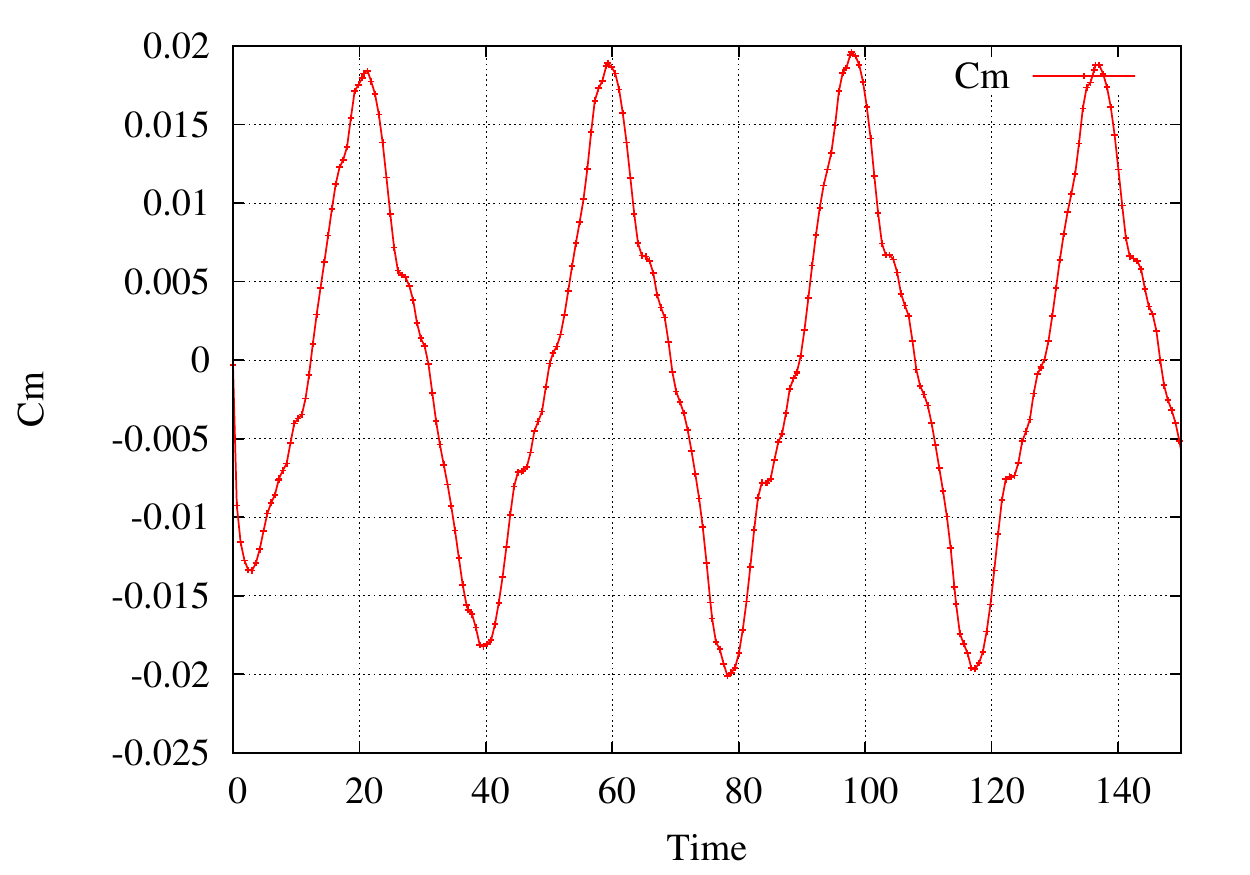} }
  \caption{Time history for normal force coefficients (a) and pitching moment coefficients (b) of the pitching NACA0012 case.}
  \label{fig:eulerNaca0012:clcm-time}
\end{figure}

The computed normal force coefficients and pitching moment coefficients, with respect to 
the angles of attack and time, respectively, 
are shown in Fig. \ref{fig:eulerNaca0012:clcm-angle} and Fig. \ref{fig:eulerNaca0012:clcm-time}. 
The numerical solutions are in good agreement with the experimental data by Landon \cite{landonNaca0012}. 
Note that while the experiment is viscous, we treat this problem as an inviscid one, 
which may lead to some difference between the numerical solution and the experimental data.

\subsection{Pitching NACA0015 Airfoil}

The laminar flow around a rapidly pitching NACA0015 airfoil is considered in this test case.
The freestream has a Mach number $\textit{Ma}_\infty = 0.2$ with specific heat ratio 
$\gamma = 1.4$. The Reynolds number based on the chord length and the freestream condition is $\textit{Re} = 10,000$.
The pitching axis is at a quarter chord length from the leading edge, and the mesh motion is 
prescribed by the pitching rate of the airfoil: 
$\omega(t) = \omega_0(1.0-\text{exp}(-4.6t/t_0))\,\,\text{rad/s}$, 
where $t_0 = c/U_\infty$ is the reference time, $\omega_0 = 0.6U_\infty/c$, 
with $c$ being the chord length and $U_\infty$ the freestream velocity. 
The flow field computed at zero angle of attack is used as the initial condition for this unsteady simulation.
  
\begin{figure}[H]
  \centering
  \subfloat[Initial Mesh]{  
    \includegraphics[trim= 2cm 1.5cm 2.2cm 2cm,clip,height = 2.5in]
    {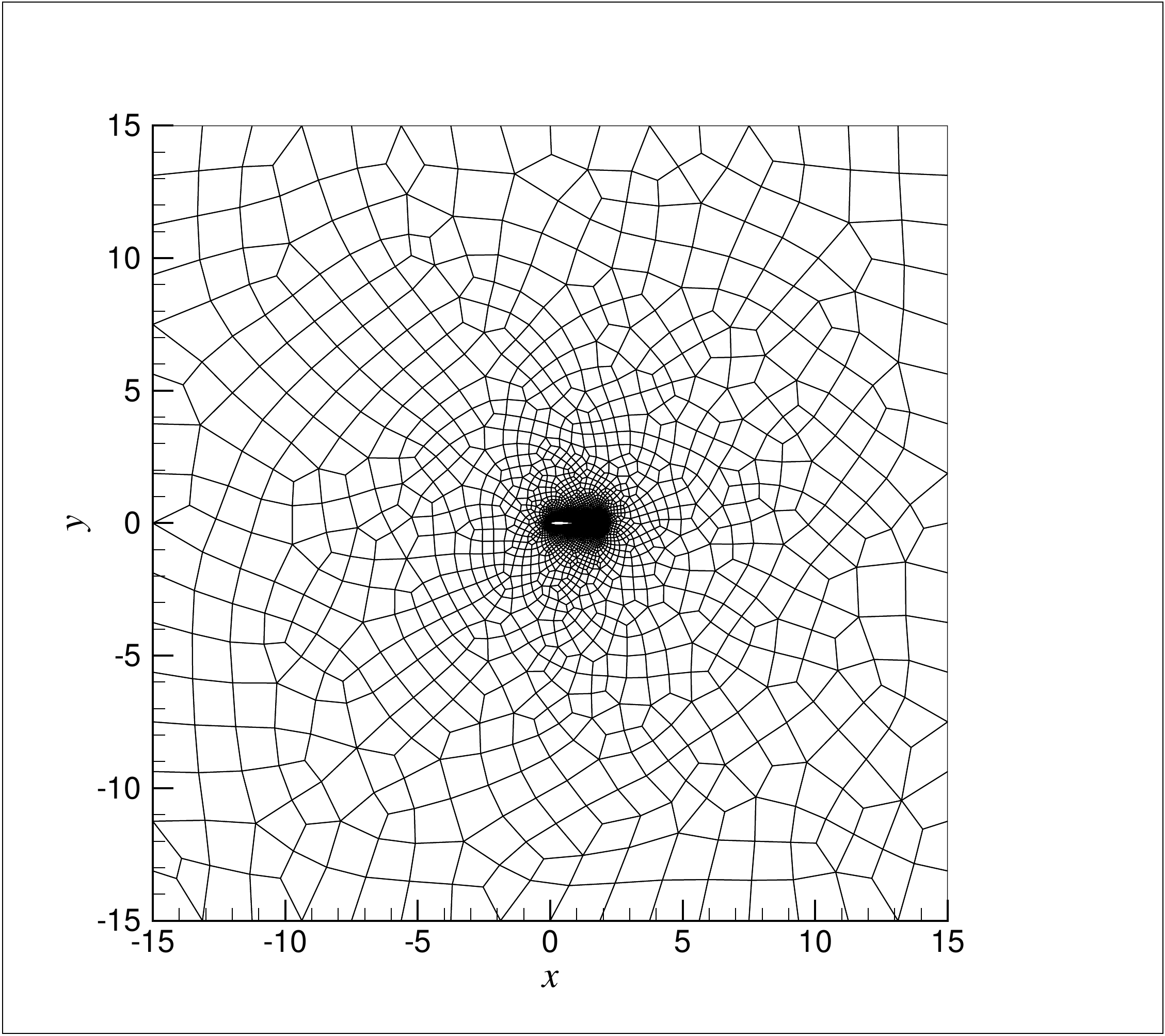} }
  \subfloat[$\alpha = 0.0$ degree]{  
    \includegraphics[trim= 2cm 1.5cm 2.2cm 2cm,clip,height = 2.5in]
    {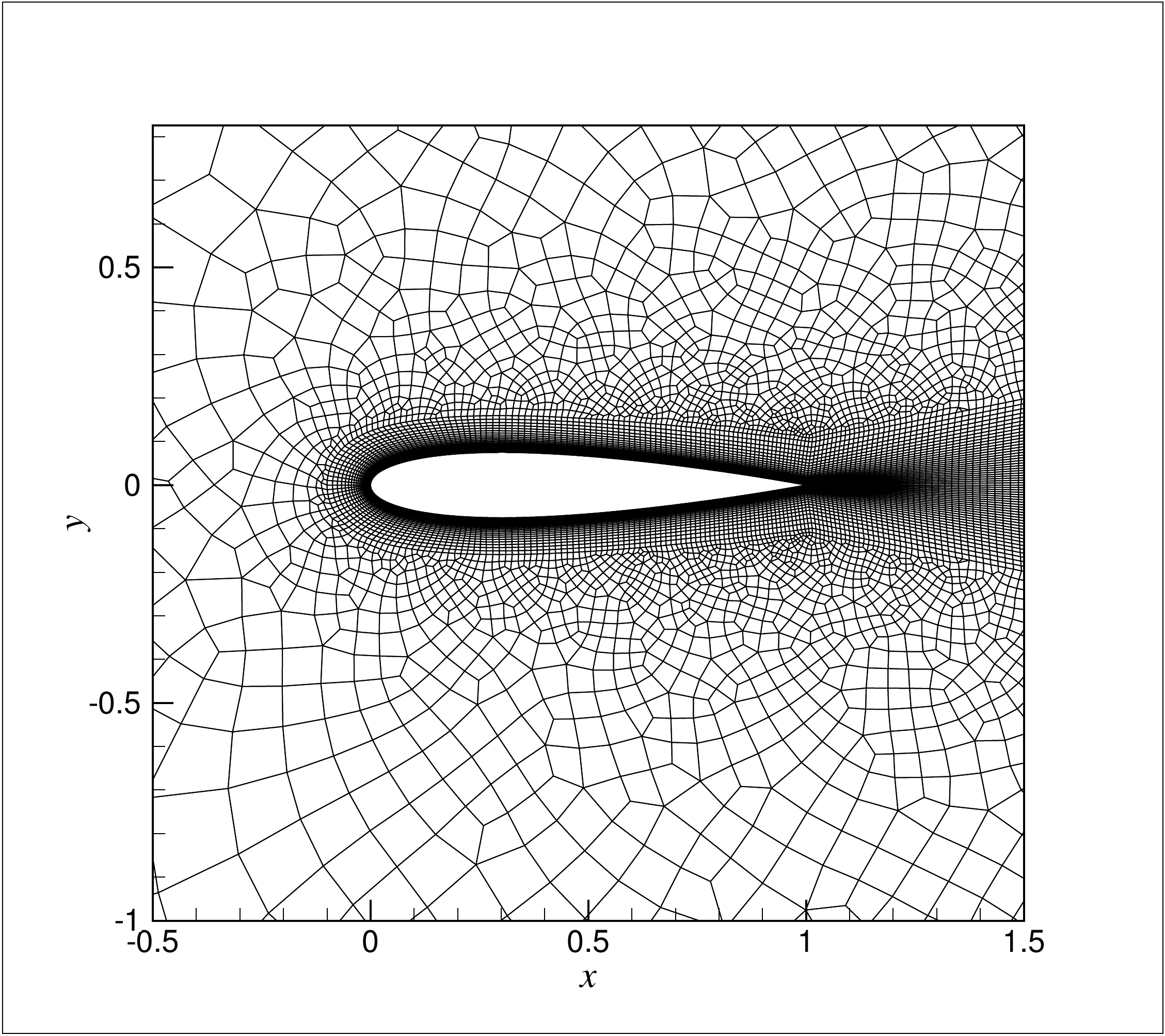} }
  \\
  \subfloat[$\alpha = 45$ degree]{  
    \includegraphics[trim= 2cm 1.5cm 2.2cm 2cm,clip,height = 2.5in]
    {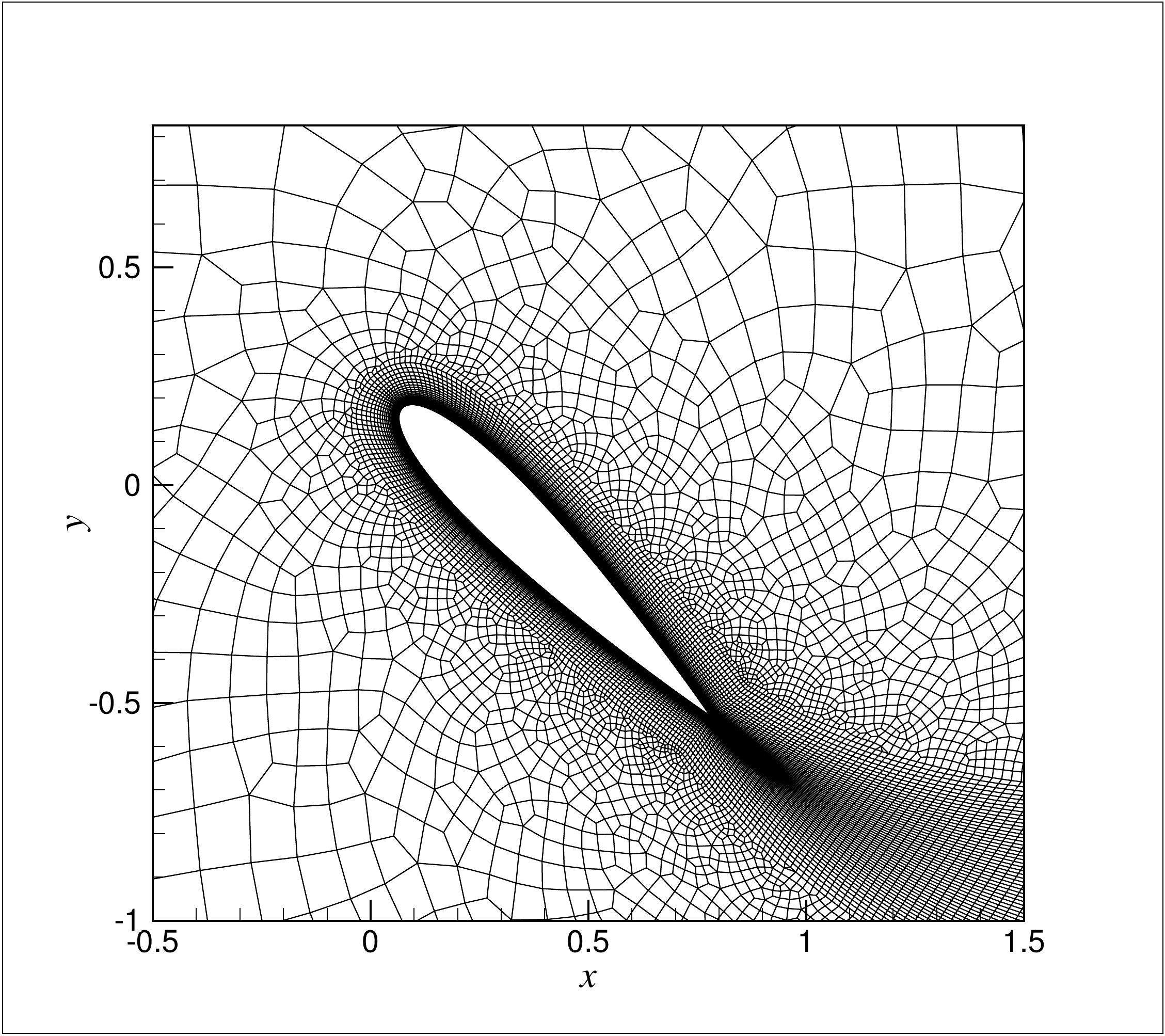} }
  \subfloat[$\alpha = 57$ degree]{  
    \includegraphics[trim= 2cm 1.5cm 2.2cm 2cm,clip,height = 2.5in]
    {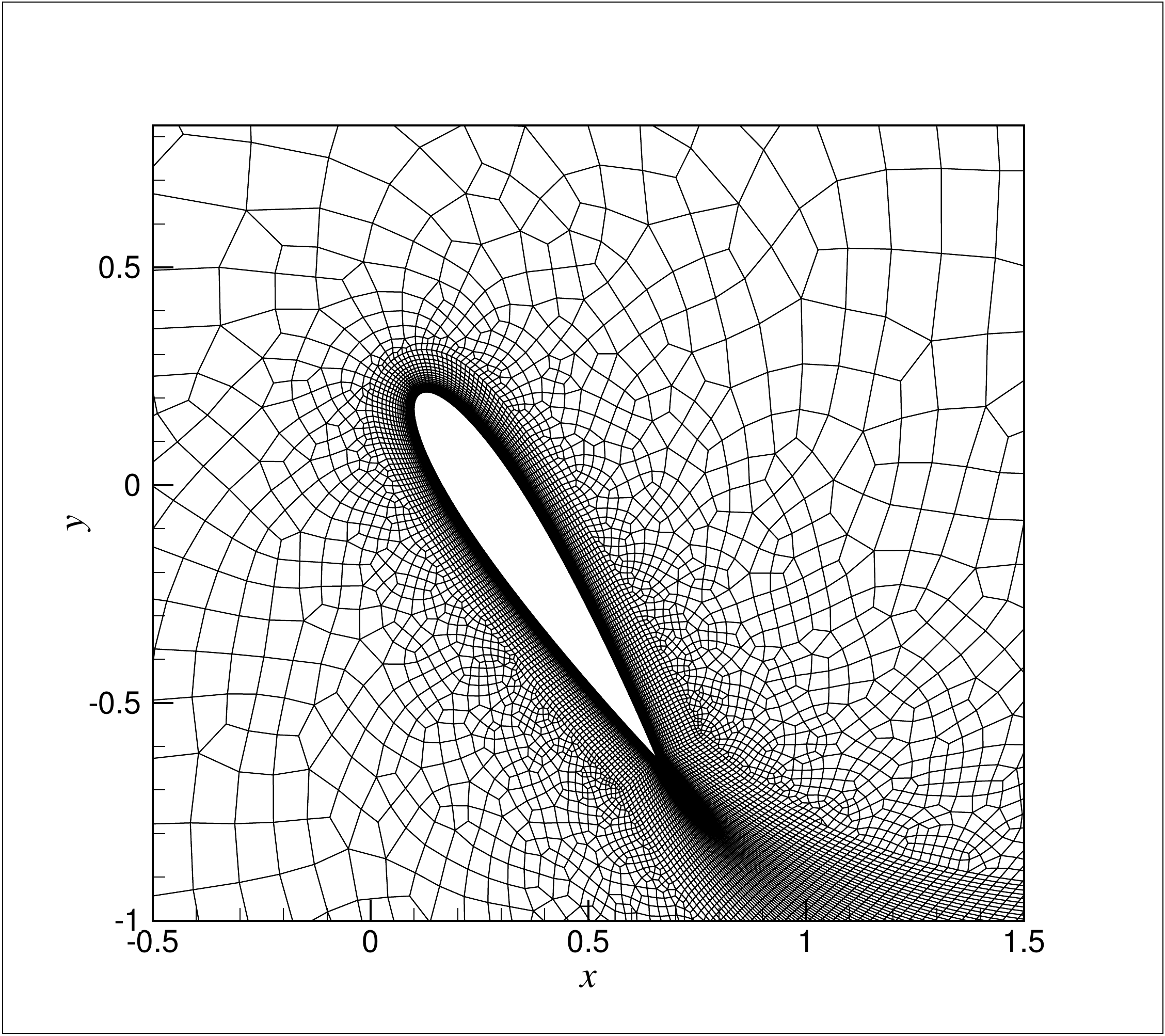} }
  \caption{Initial mesh  and mesh at three different angles of attack for the pitching NACA0015 case.}
  \label{fig:nsNaca0015:mesh}
\end{figure}

The computational domain is $(-15,15)\times (-15,15)$, with the airfoil located at the center.
The initial mesh and the zoomed-in mesh near the airfoil at three 
different angles of attack are shown in Fig. \ref{fig:nsNaca0015:mesh}.
As usual, the RBF method is used to compute the movement of the interior mesh nodes, i.e., 
the nodes on the airfoil move according to the above variation of the angle of attack,
while those at far-field are kept static.
It can be seen that the mesh quality near the wall is well maintained, even at large angle of attack.
Here we have two set of curved grids. For the coarse mesh, the minimum and maximum wall normal
spacings are 0.005 and 0.008, respectively, with 180 nodes on the airfoil, 
while for the fine mesh, the minimum and maximum normal spacings are 0.00037 and 0.0006, with 200 nodes on the airfoil.

\begin{figure}[H]
  \centering
  \subfloat[$\alpha = 45$ degree]{  
    \includegraphics[trim= 1.5cm 1.5cm 0.1cm 2cm,clip,height = 2.5in]
    {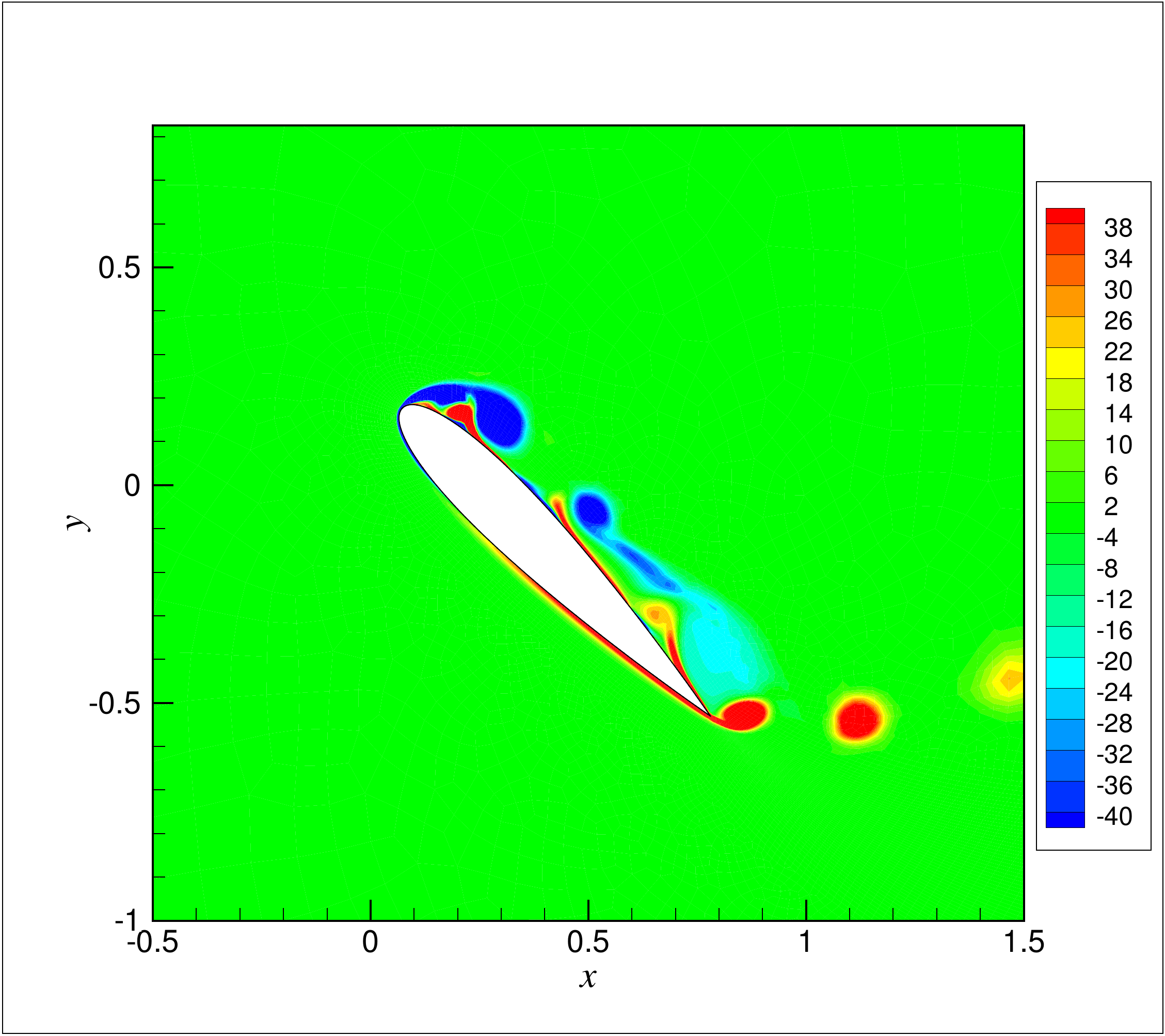} } 
  \\ 
  \subfloat[$\alpha = 50$ degree]{  
    \includegraphics[trim= 1.5cm 1.5cm 0.1cm 2cm,clip,height = 2.5in]
    {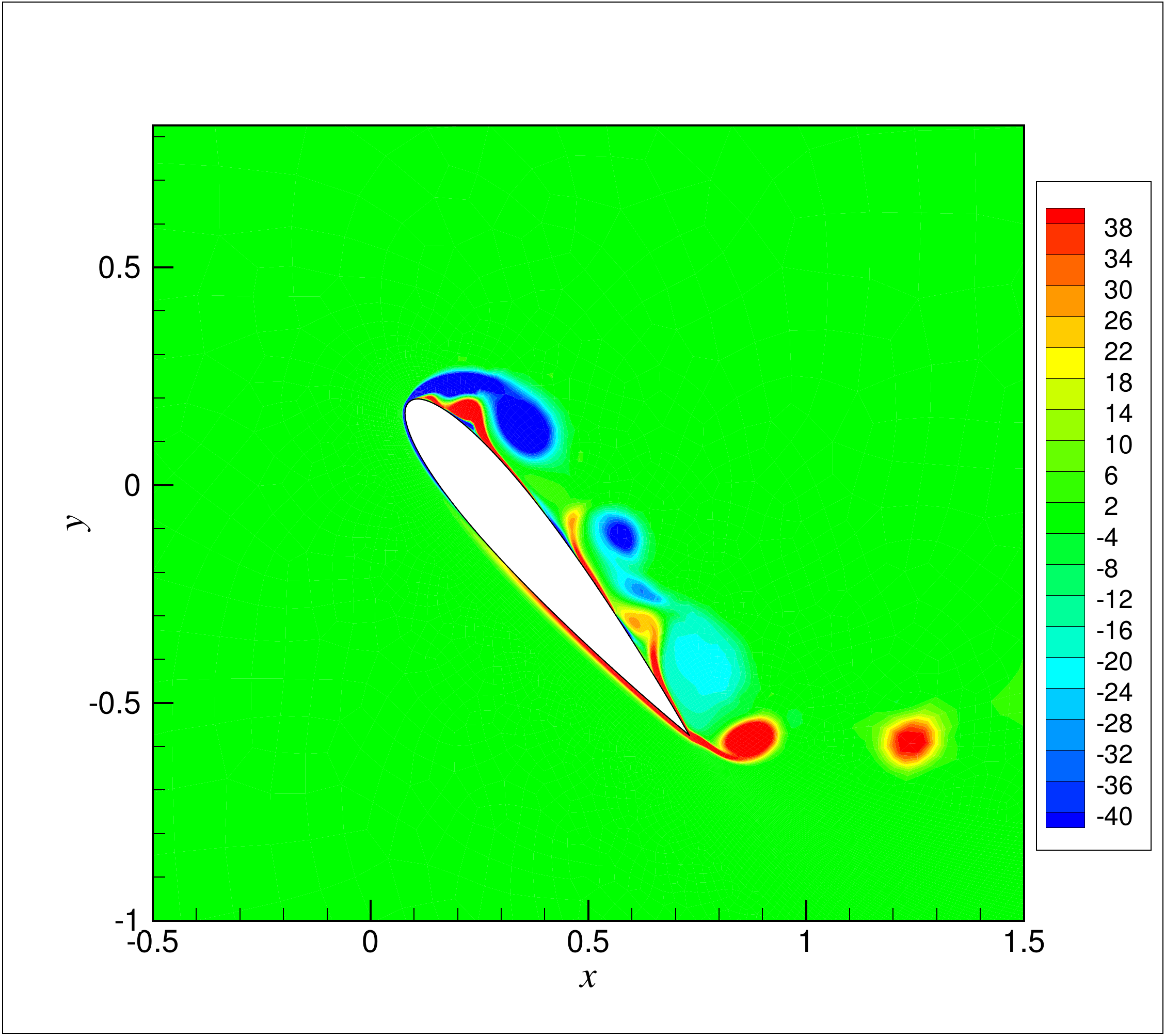} }
  \\
  \subfloat[$\alpha = 57$ degree]{  
    \includegraphics[trim= 1.5cm 1.5cm 0.1cm 2cm,clip,height = 2.5in]
    {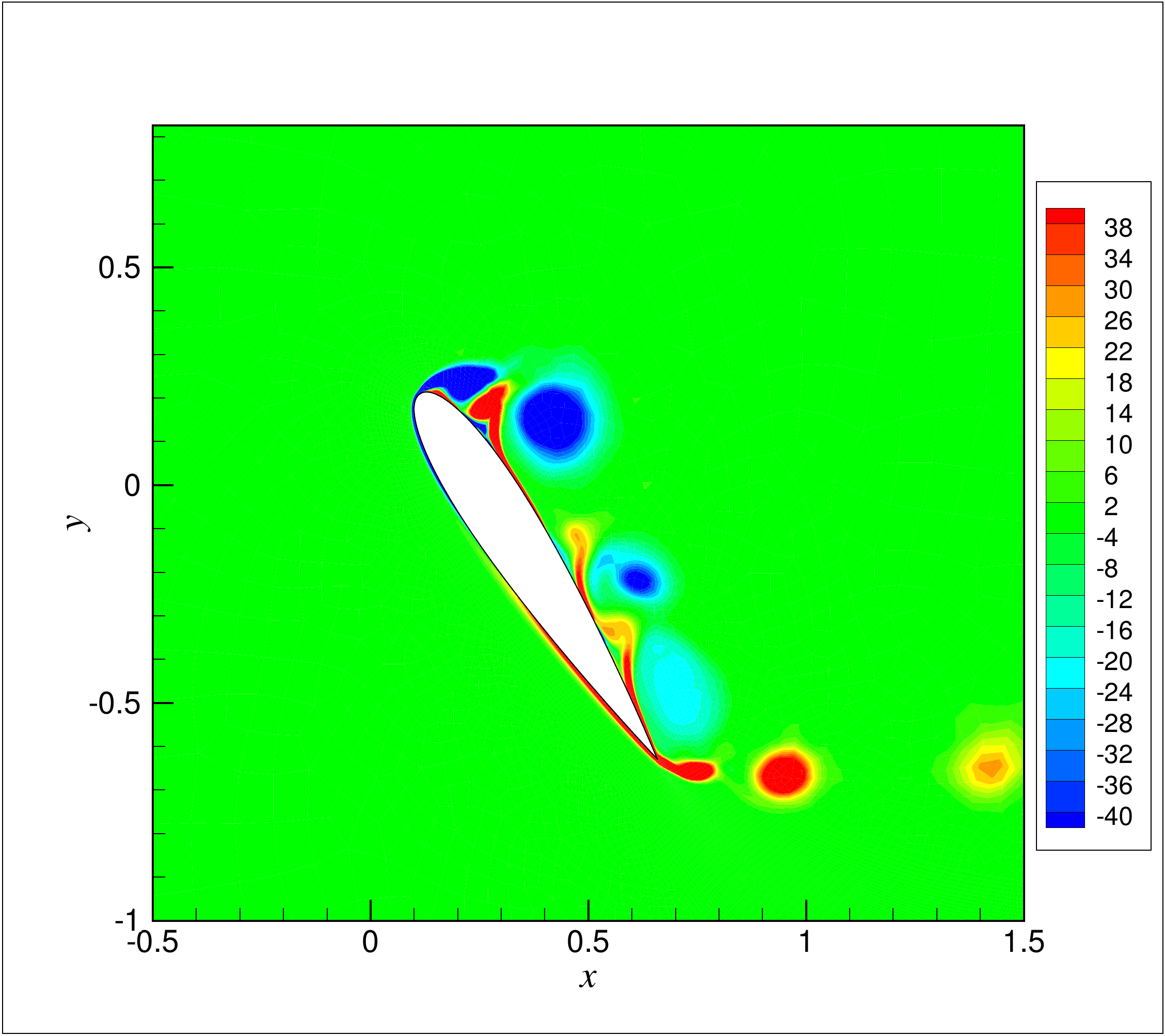} }
  \caption{Normalized vorticity contour at three different angles of attack for the pitching NACA0015 case. }
  \label{fig:nsNaca0015:vorticity}
\end{figure}

In Fig. \ref{fig:nsNaca0015:vorticity}, the instantaneous normalized vorticities 
$\omega_z c/U_\infty$ at three different angles of attack are presented. 
It can be seen the vorticities near the wall of the airfoil formed and transported 
towards downstream. 
The computed lift coefficients and drag coefficients are compared with the simulation data 
from Visbal et al. \cite{visbal1989naca0015} and Ren and Xu et al. \cite{renxu2016ale}.
From the comparison we can see that the computed results with coarse mesh are in good 
agreement with the reference data, while those on the fine mesh have relatively large 
discrepancies with the references, especially for the lift coefficients.  
In general, fine mesh should better capture the flow structures due to its finer resolution.
While the reason for the occurrence of this phenomenon is not quite clear and is still under 
investigation, there could be a number of factors that might contribute, for example, as
pointed out in \cite{renxu2016ale}, how to compute the mesh velocity at the cell interface
has a significant effect on the numerical solution, and a variable mesh velocity along cell
interface is necessary for a high order method, after comparing the results from a piecewise 
constant grid velocity and a variable mesh velocity  \cite{renxu2016ale}. 
In the current work, a quadratic 
interpolation from the three nodes of an interface (two end nodes and one midpoint) is used
to obtain the variable mesh velocities.

\begin{figure}[H]
  \centering
  \includegraphics[height =4.0in]
    {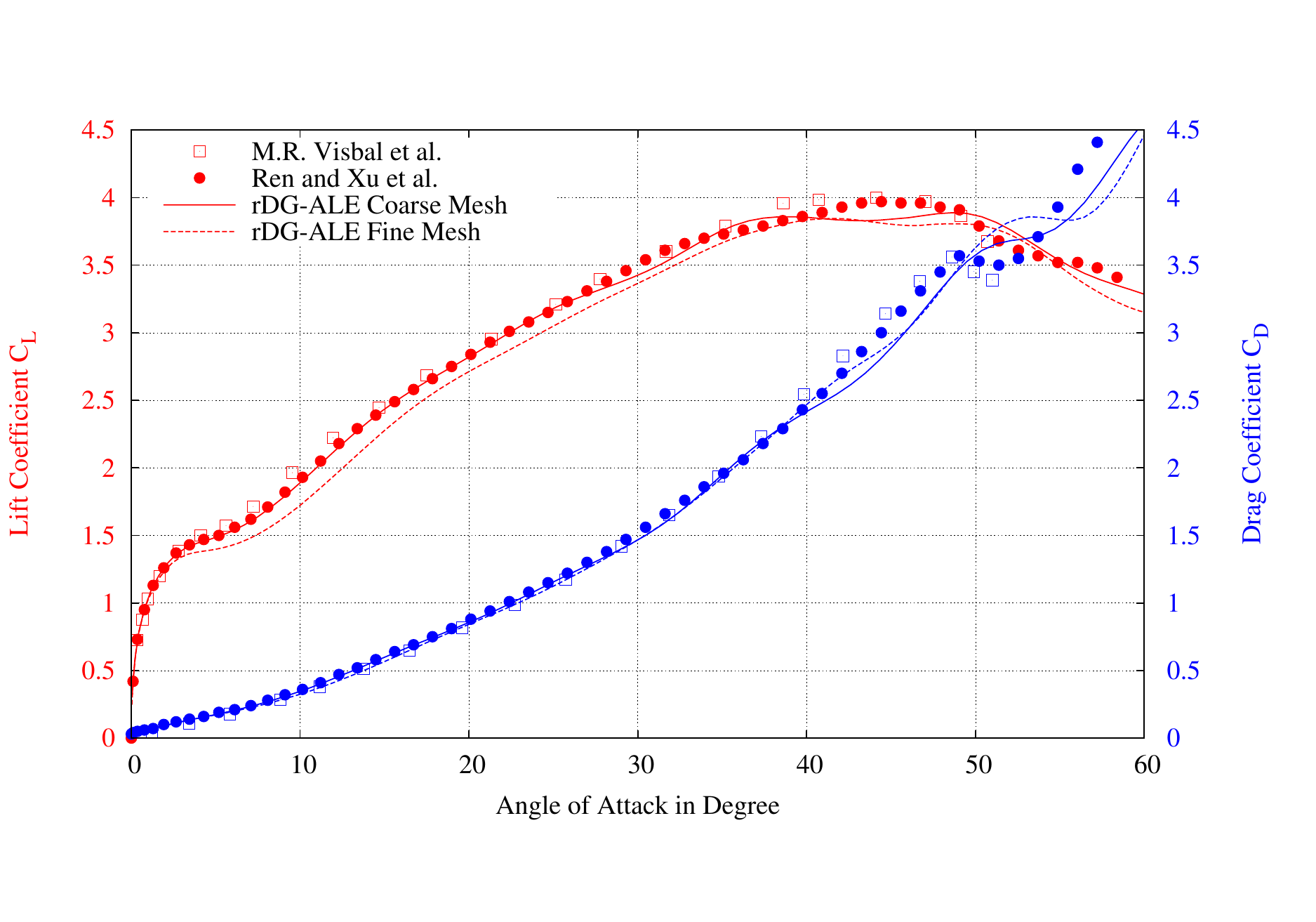}
  \caption{Lift coefficients and drag coefficients versus angles of attack  for the pitching NACA0015 case.}
  \label{fig:nsNaca0015:clcd}
\end{figure}

\subsection{Plunging SD7003 Airfoil}

The flow over a moving SD7003 airfoil has been investigated both experimentally 
and numerically in the literature. 
McGowan et al. \cite{mcgowan2008sd7003,mcgowan2011sd7003} conducted a set of experiments 
on the purely plunging or pitching SD7003 airfoil, using particle image velocimetry (PIV) 
in a water tunnel, besides, the numerical solutions from both CFL3D and an immersed boundary 
method are compared with the experimental data. 
Visbal et al. \cite{visbal2009sd7003} performed the computations in which the grid is moved in a rigid fashion, 
as opposed to the smoothing or interpolation approaches, e.g., the RBF method. 

In this work, the flow over the high-frequency plunging SD7003 airfoil is simulated using
the rDG-ALE method. 
The Reynolds number based on the chord length and the freestream velocity is
$\textit{Re}=10,000$, and the specific heat ratio $\gamma = 5/3$. 
The airfoil is set at a static angle of attack $\alpha_0 = 4$ degree.
The plunging motion is prescribed as $h(t) = h_0\text{sin}(2kU_\infty t/c)$ where $h_0 = 0.05c$ 
is the plunging amplitude and $k=3.93$ is the reduced frequency, 
with $c$ the chord length and $U_\infty$ the freestream velocity. 
Although the motion-induced angle of attack could be as high as $21.5$ degree which leads to 
unsteady separation and the generation of dynamic-stall-like vortices at the leading edge,
the transition effects are observed to be minor for this low Reynolds number 
($\textit{Re}=10,000$) \cite{visbal2009sd7003}.
To show the compressibility effect for this test case, we choose two Mach numbers,
$\textit{Ma}=0.2$ and 0.05, as in Ren and Xu's paper \cite{renxu2016ale}.

\begin{figure}[H]
  \centering
  \subfloat[Initial mesh]{  
    \includegraphics[trim= 2cm 1.5cm 2.2cm 2cm,clip,height = 3.2in]
    {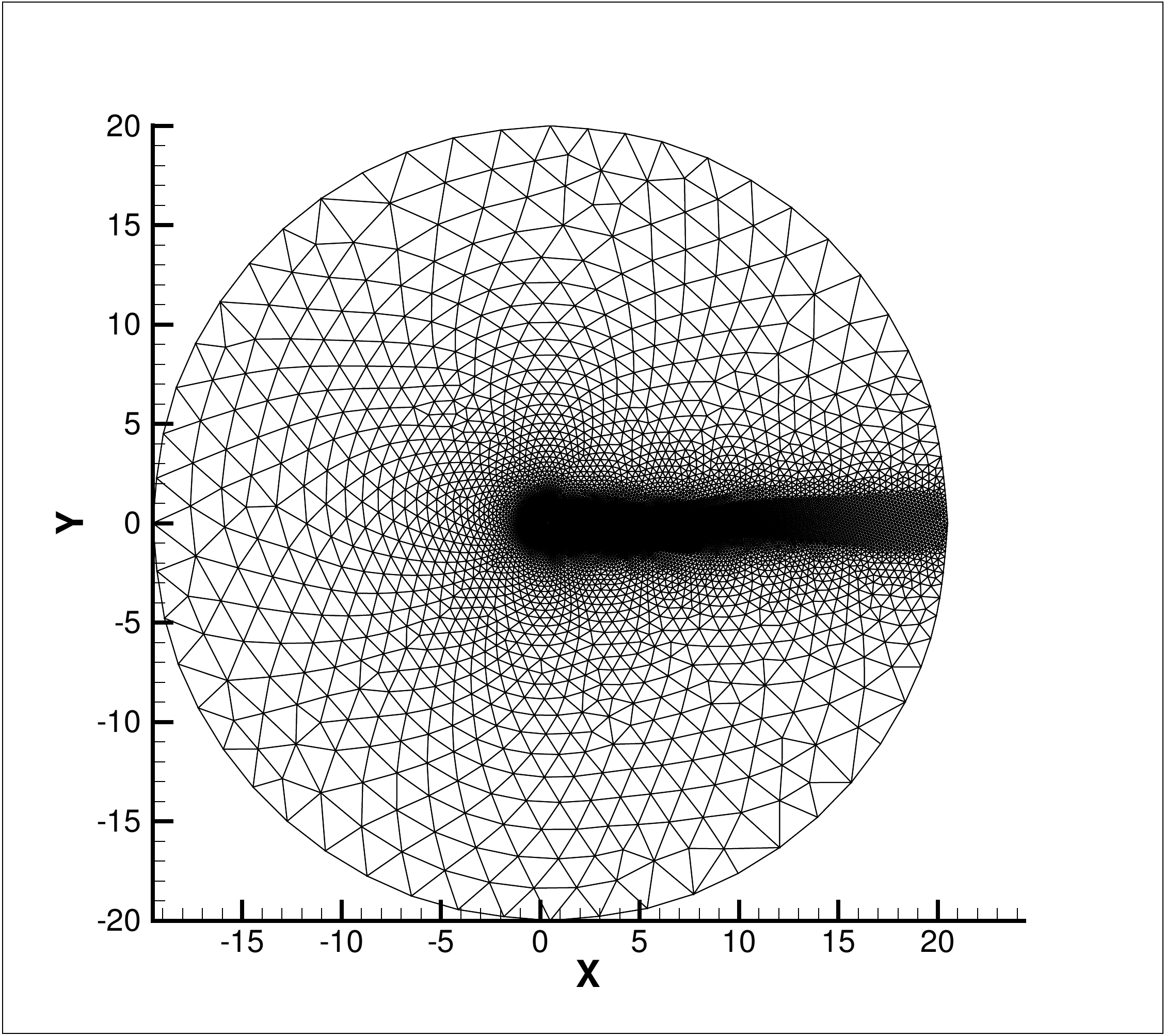} } 
  \\ 
  \subfloat[$h = 0.0$]{  
    \includegraphics[trim= 2cm 1.5cm 2.2cm 2cm,clip,height = 2.5in]
    {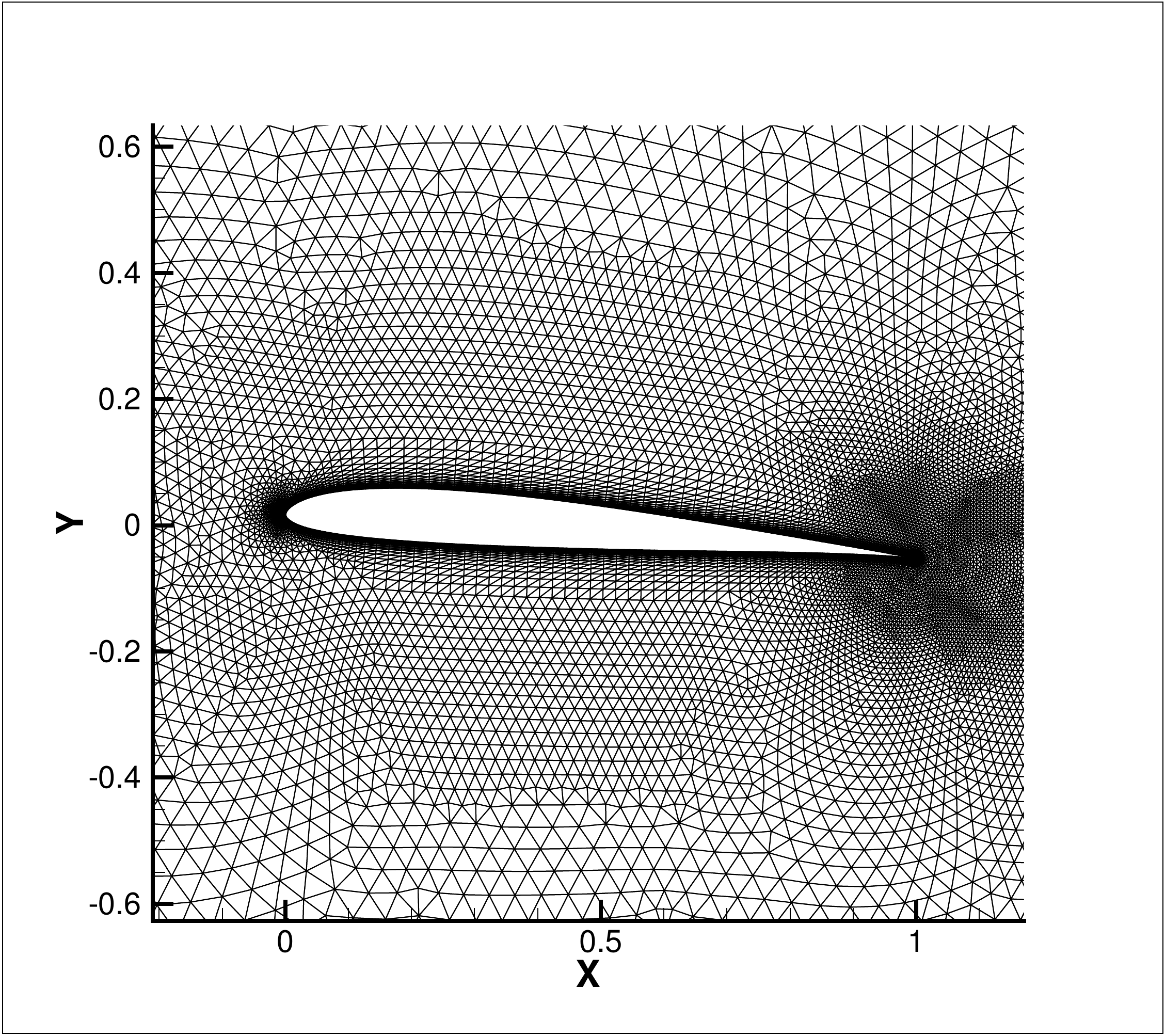} }
  \subfloat[$h = h_0$]{  
    \includegraphics[trim= 2cm 1.5cm 2.2cm 2cm,clip,height = 2.5in]
    {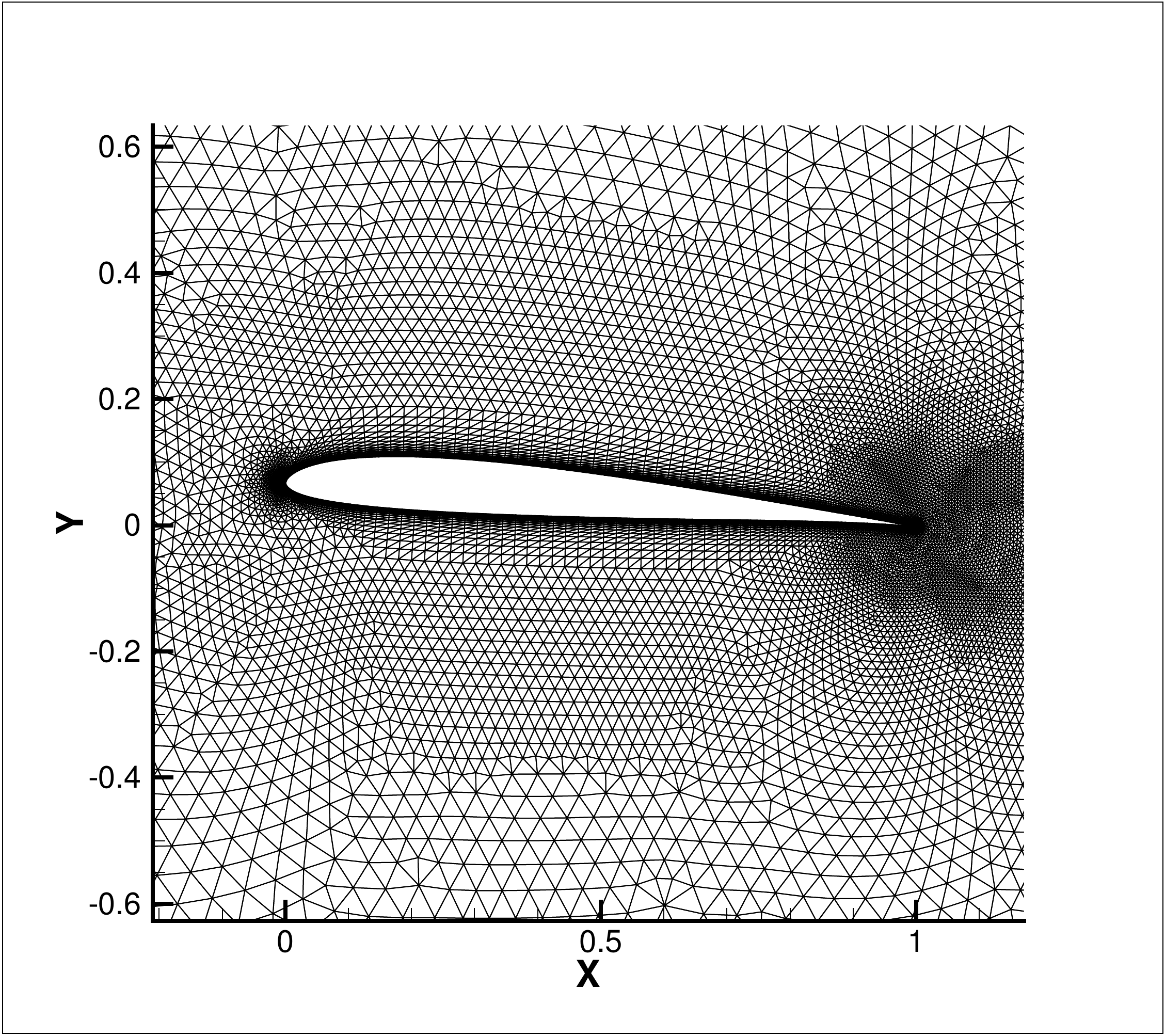} }
  \caption{Zoomed-in mesh near the airfoil of the plunging SD7003 case. left:initial mesh; right: mesh at maximum angle of attack}
  \label{fig:nsSD7003:mesh}
\end{figure}

\begin{figure}[H]
  \centering
  \subfloat[]{  
    \includegraphics[trim= 0cm 0cm 0.5cm 0cm,clip,height = 2.5in]
    {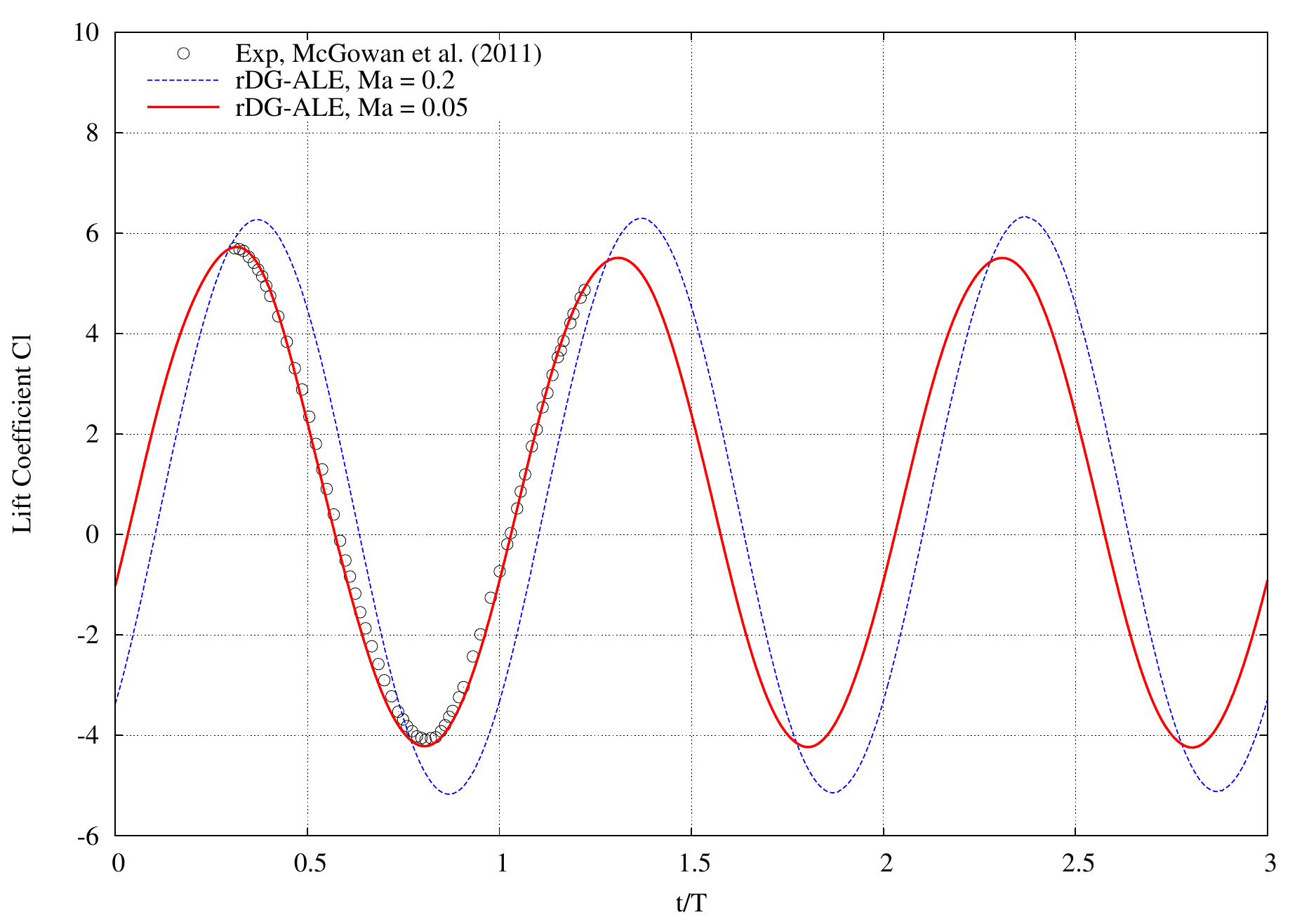} }
  \\
  \subfloat[]{  
    \includegraphics[trim= 0cm 0cm 0.5cm 0cm,clip,height = 2.5in]
    {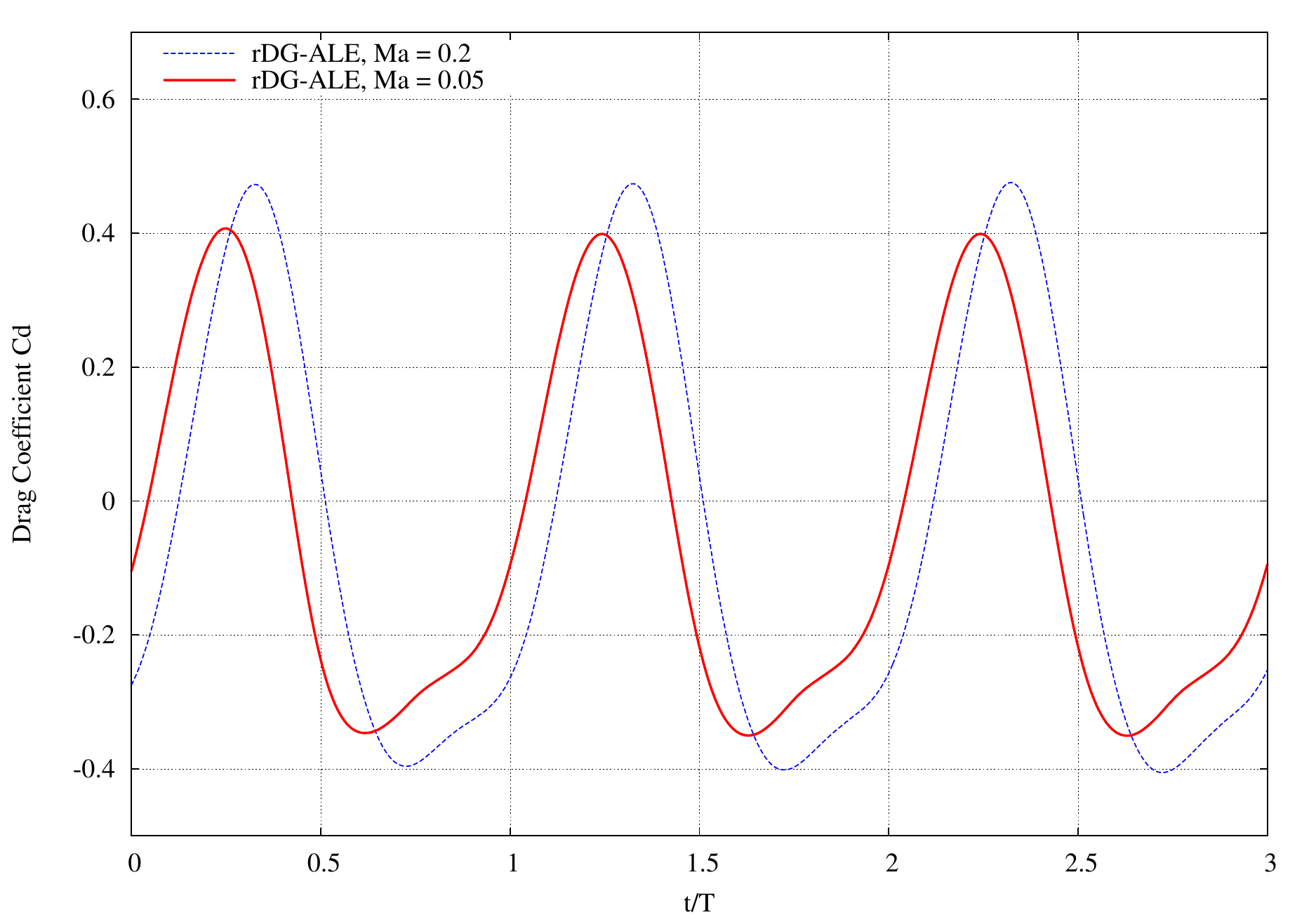} }
  \caption{Lift coefficients (a) and drag coefficients (b) versus normalized time  for the plunging SD7003 case.}
  \label{fig:nsSD7003:clcd}
\end{figure}

\begin{figure}[H]
  \centering
  \subfloat[Experiment]{
    \begin{minipage}{0.5\linewidth}
      \centering  
      \includegraphics[trim= 0cm 0cm 0cm 0cm,clip,width = 2.6in]
      {./fig/nsSD7003/expLegend.png}
      \includegraphics[trim= 0cm 0cm 0cm 0cm,clip,width = 3.1in,frame]
      {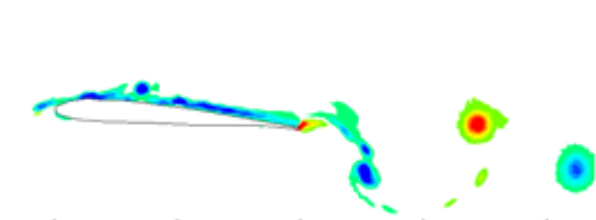}
    \end{minipage}  
  }
  \\  
  \subfloat[Experiment]{
    \hspace{0.7in}
    \includegraphics[trim= 0cm 0cm 0cm 0cm,clip,width = 3.1in]
    {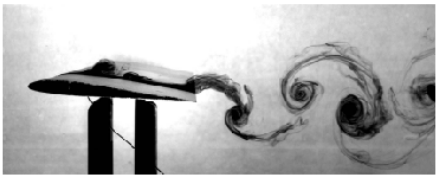}
  }
  \\
  \subfloat[rDG-ALE, Ma = 0.2]{
    \begin{minipage}{0.5\linewidth}
      \centering  
      \includegraphics[trim= 2cm 9.5cm 2.2cm 8.5cm,clip,width = 2.8in]
      {./fig/nsSD7003/vorticityT20_mach0p05-eps-converted-to.pdf}
      \includegraphics[trim= 2cm 1.5cm 2.2cm 10.5cm,clip,width = 3.1in]
      {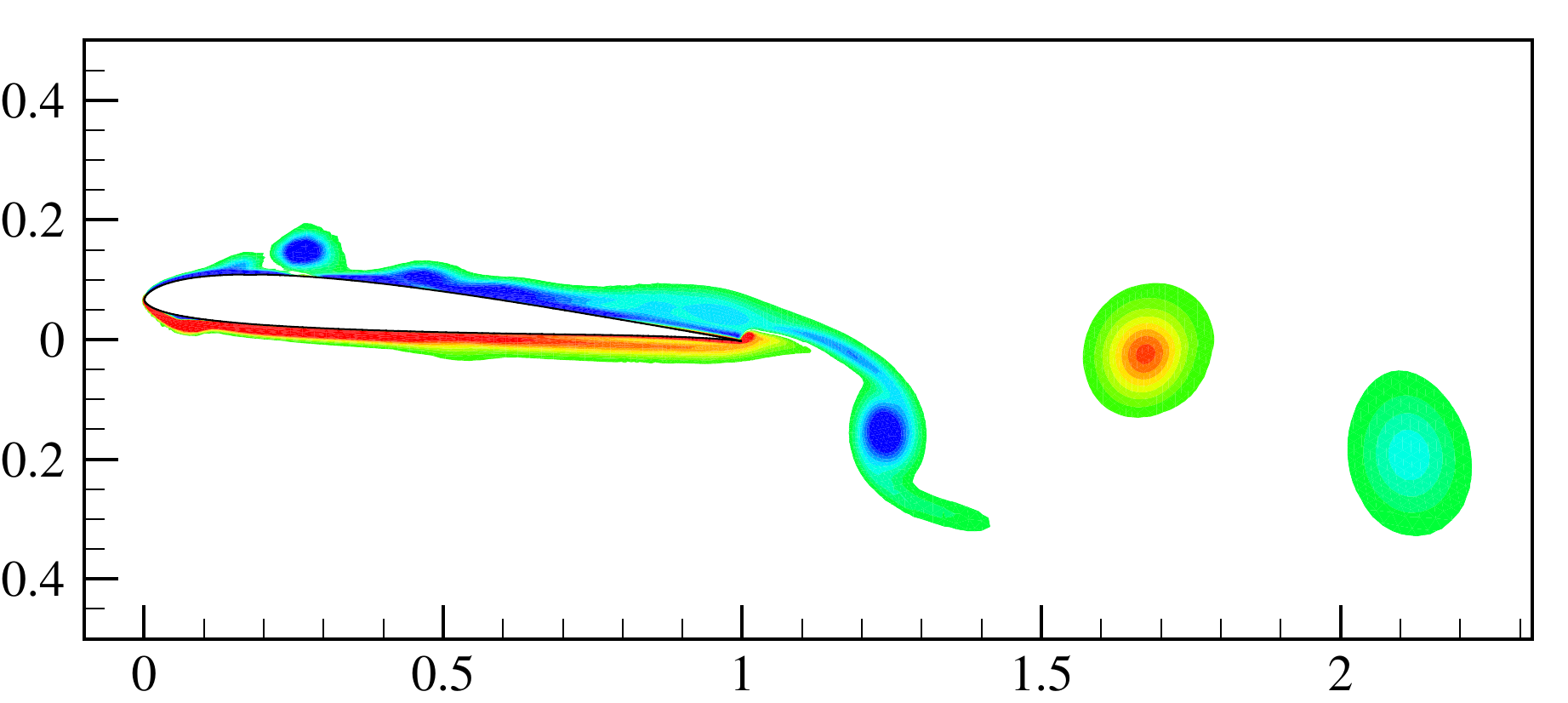}
    \end{minipage}  
  }
  \\
  \subfloat[rDG-ALE, Ma = 0.05]{
    \hspace{0.7in}
    \includegraphics[trim= 2cm 1.5cm 2.2cm 10.5cm,clip,width = 3.1in]
    {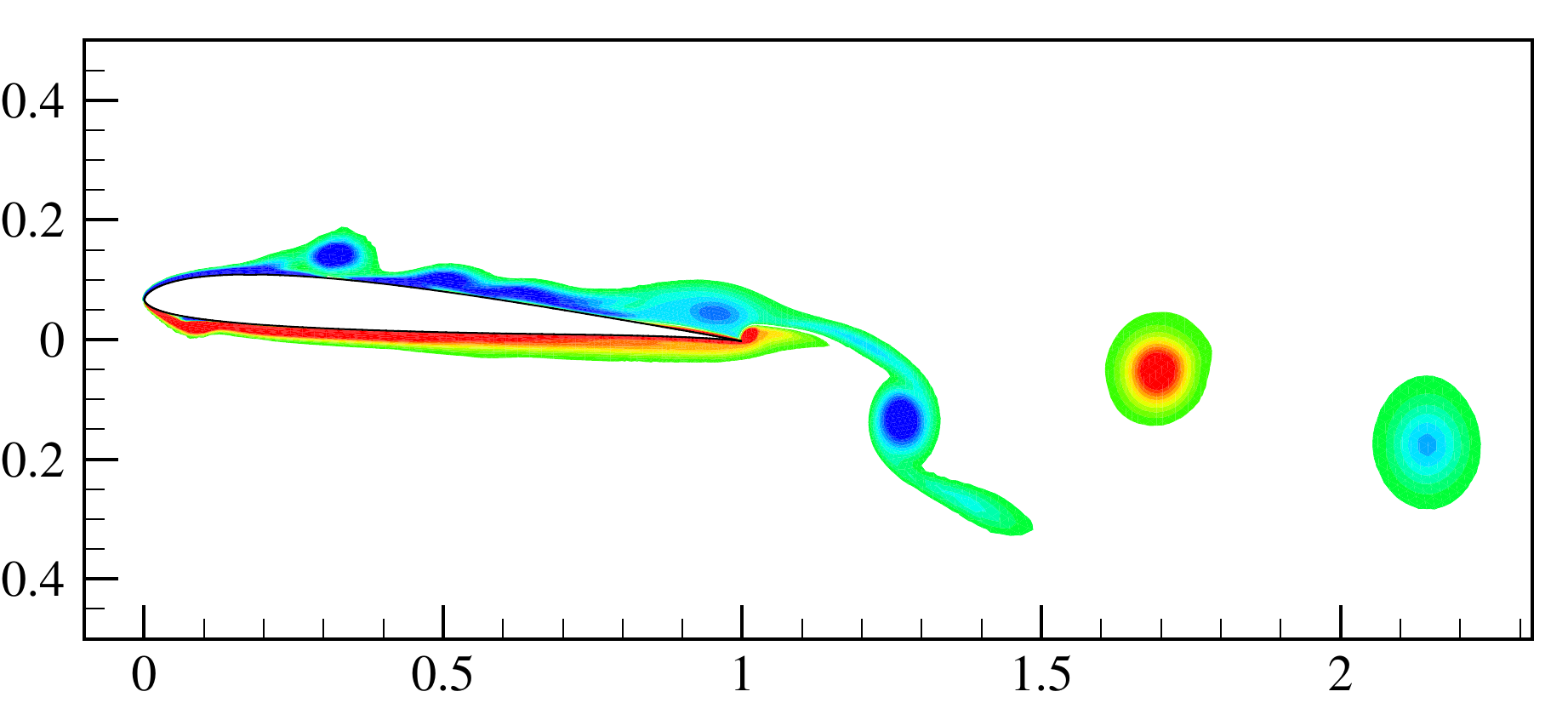}
  }
  \caption{Comparison of normalized vorticities with experiment at $t=1/4 T$  for the plunging SD7003 case.}
  \label{fig:nsSD7003:vorticity:0.25T}
\end{figure}

\begin{figure}[H]
  \centering
  \subfloat[Experiment]{
    \begin{minipage}{0.5\linewidth}
    \centering
      \includegraphics[trim= 0cm 0cm 0cm 0cm,clip,width = 2.6in]
      {./fig/nsSD7003/expLegend.png}
      \includegraphics[trim= 0cm 0cm 0cm 0cm,clip,width = 3.1in,frame]
      {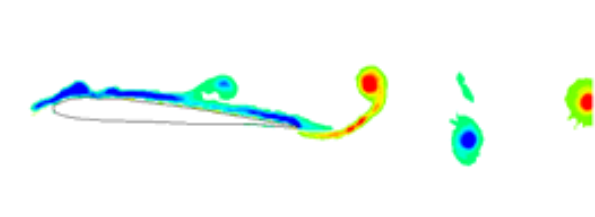}
    \end{minipage}
  }
  \\
  \subfloat[Experiment]{
    \hspace{0.7in}
    \includegraphics[trim= 0cm 0cm 0cm 0cm,clip,width = 3.1in]
    {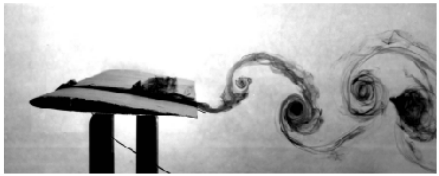}
  }
  \\
  \subfloat[rDG-ALE, Ma = 0.2]{
    \begin{minipage}{0.5\linewidth}
      \centering  
      \includegraphics[trim= 2cm 9.5cm 2.2cm 8.5cm,clip,width = 2.8in]
      {./fig/nsSD7003/vorticityT20_mach0p05-eps-converted-to.pdf}
      \includegraphics[trim= 2cm 1.5cm 2.2cm 10.5cm,clip,width = 3.1in]
      {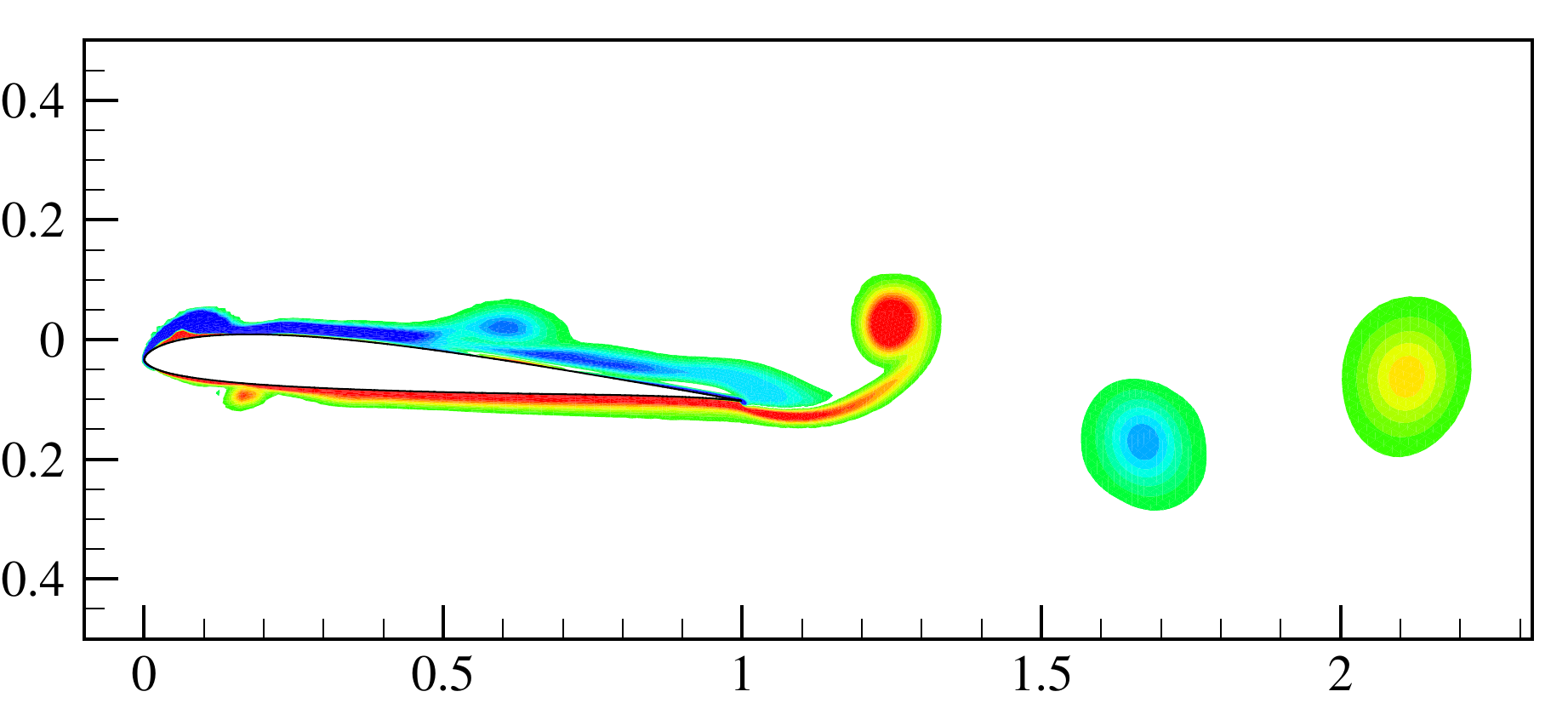}
    \end{minipage}
  }
  \\
  \subfloat[rDG-ALE, Ma = 0.05]{
    \hspace{0.7in}
    \includegraphics[trim= 2cm 1.5cm 2.2cm 10.5cm,clip,width = 3.1in]
    {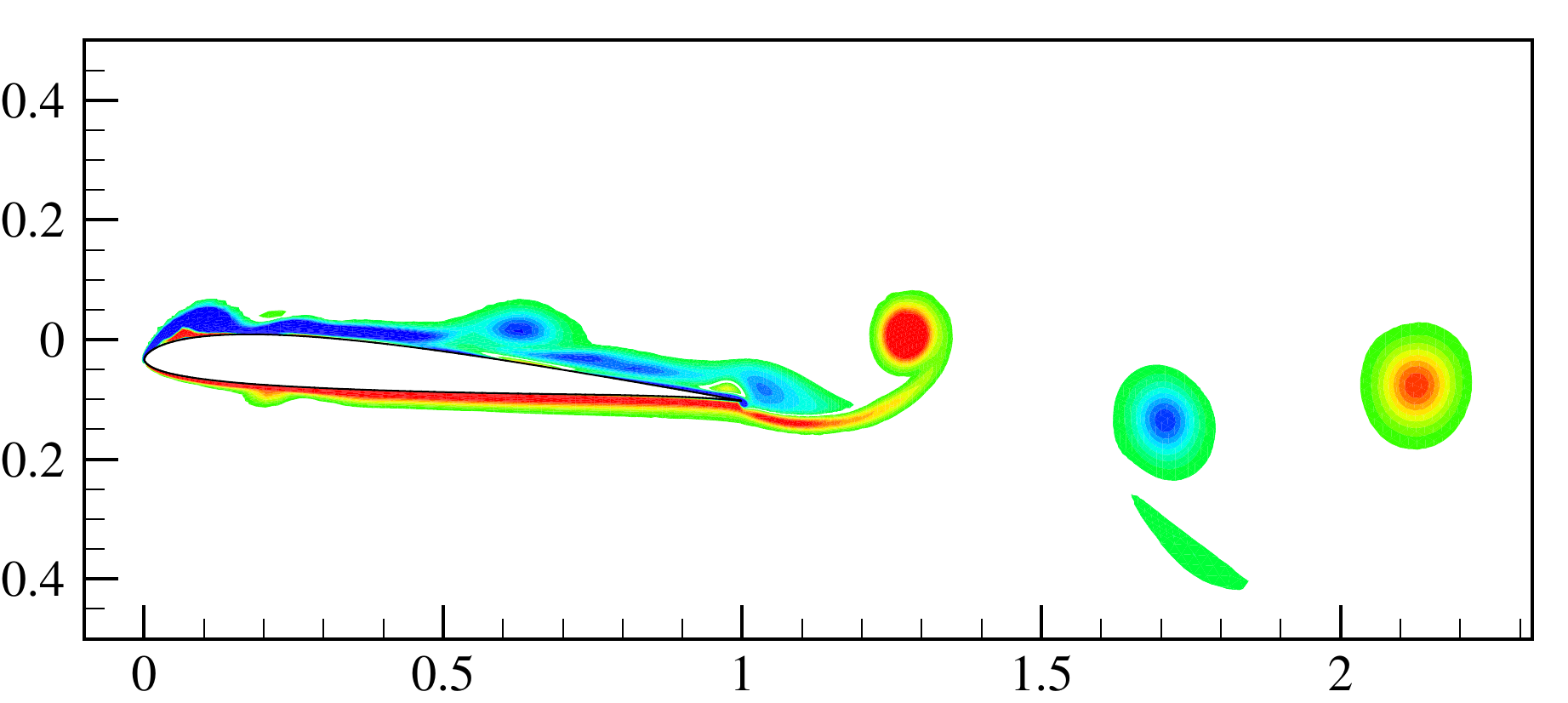}
  }
  \caption{Comparison of normalized vorticities with experiment at $t=3/4 T$  for the plunging SD7003 case.}
  \label{fig:nsSD7003:vorticity:0.75T}
\end{figure}

The computational domain is a circle region with radius $R=20$. 
Fig. \ref{fig:nsSD7003:mesh} shows the initial mesh and zoomed-in mesh.
The number of nodes on the airfoil is 200 and the wall normal spacing is approximately 0.0006. 
The time-dependent lift coefficients and drag coefficients are compared with the 
experimental data from McGowan et al. \cite{mcgowan2008sd7003,mcgowan2011sd7003}., 
and shown in Fig. \ref{fig:nsSD7003:clcd}. 
We can see that at $\textit{Ma}=0.05$ the lift coefficient has an excellent agreement with 
the experimental data, while a phase lag is observed for both the lift and drag coefficients at $\textit{Ma}=0.2$. 
The computed vorticities are also compared with the experimental data, 
shown in Fig. \ref{fig:nsSD7003:vorticity:0.25T} and Fig. \ref{fig:nsSD7003:vorticity:0.75T}, respectively, for two time instances. 
We can see they agree well with each other.

\begin{figure}[H]
  \centering
  \subfloat[Experiment]{
    \begin{minipage}{0.5\linewidth}
      \centering  
      \includegraphics[trim= 0cm 0cm 0cm 0cm,clip,width = 2.6in,frame]
      {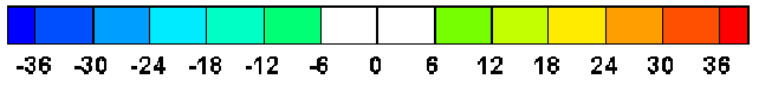}
      \includegraphics[trim= 0cm 0cm 0cm 0cm,clip,width = 3.1in,frame]
      {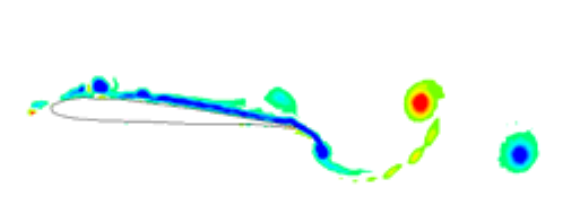}
    \end{minipage}  
  }
  \\
  \subfloat[rDG-ALE, Ma = 0.2]{
    \begin{minipage}{0.5\linewidth}
      \centering  
      \includegraphics[trim= 2cm 9.5cm 2.2cm 8.5cm,clip,width = 2.8in]
      {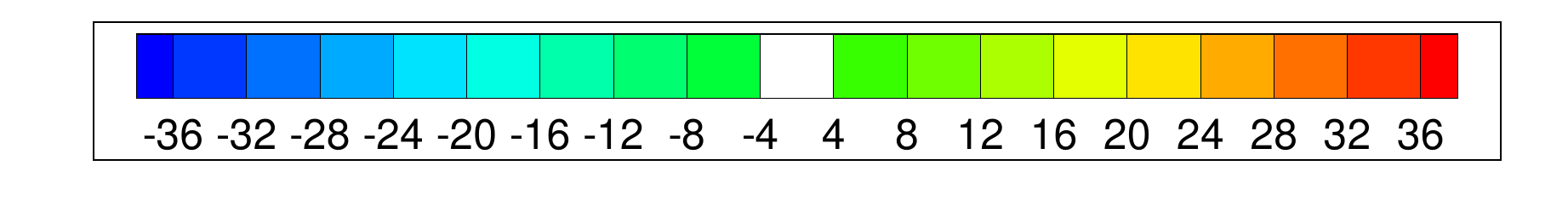}
      \includegraphics[trim= 2cm 1.5cm 2.2cm 10.5cm,clip,width = 3.1in]
      {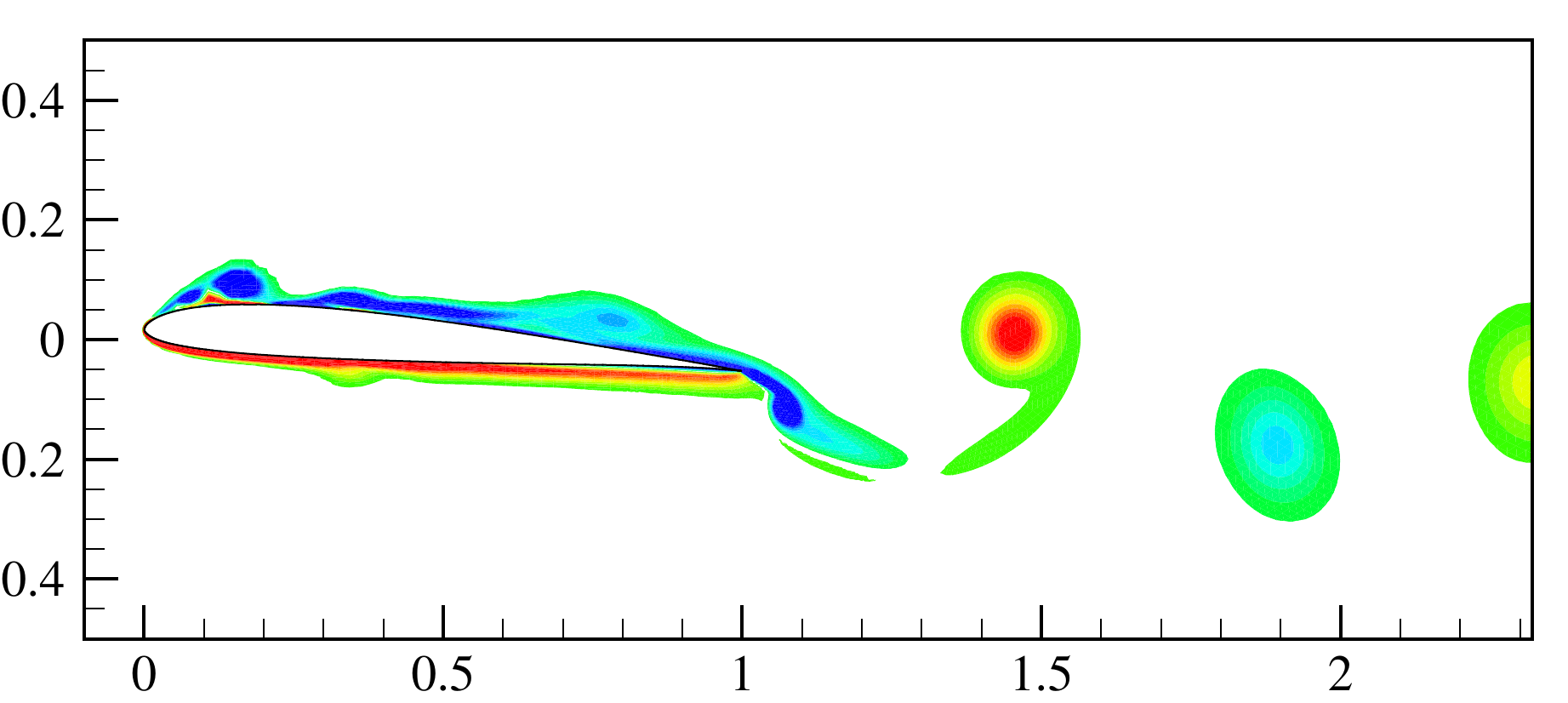}
    \end{minipage}  
  }
  \\
  \subfloat[rDG-ALE, Ma = 0.05]{
    \hspace{0.7in}
    \includegraphics[trim= 2cm 1.5cm 2.2cm 10.5cm,clip,width = 3.1in]
    {./fig/nsSD7003/vorticityT20_mach0p05-eps-converted-to.pdf}
  }
  \caption{Comparison of normalized vorticities with experiment at $t=0 T$  for the plunging SD7003 case.}
  \label{fig:nsSD7003:vorticity:0T}
\end{figure}

\begin{figure}[H]
  \centering
  \subfloat[Experiment]{
    \begin{minipage}{0.5\linewidth}
      \centering  
      \includegraphics[trim= 0cm 0cm 0cm 0cm,clip,width = 2.6in]
      {./fig/nsSD7003/expLegend.png}
      \includegraphics[trim= 0cm 0cm 0cm 0cm,clip,width = 3.1in,frame]
      {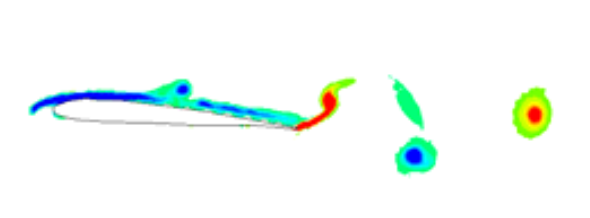}
    \end{minipage}  
  }
  \\
  \subfloat[rDG-ALE, Ma = 0.2]{
    \begin{minipage}{0.5\linewidth}
      \centering  
      \includegraphics[trim= 2cm 9.5cm 2.2cm 8.5cm,clip,width = 2.8in]
      {./fig/nsSD7003/vorticityT20_mach0p05-eps-converted-to.pdf}
      \includegraphics[trim= 2cm 1.5cm 2.2cm 10.5cm,clip,width = 3.1in]
      {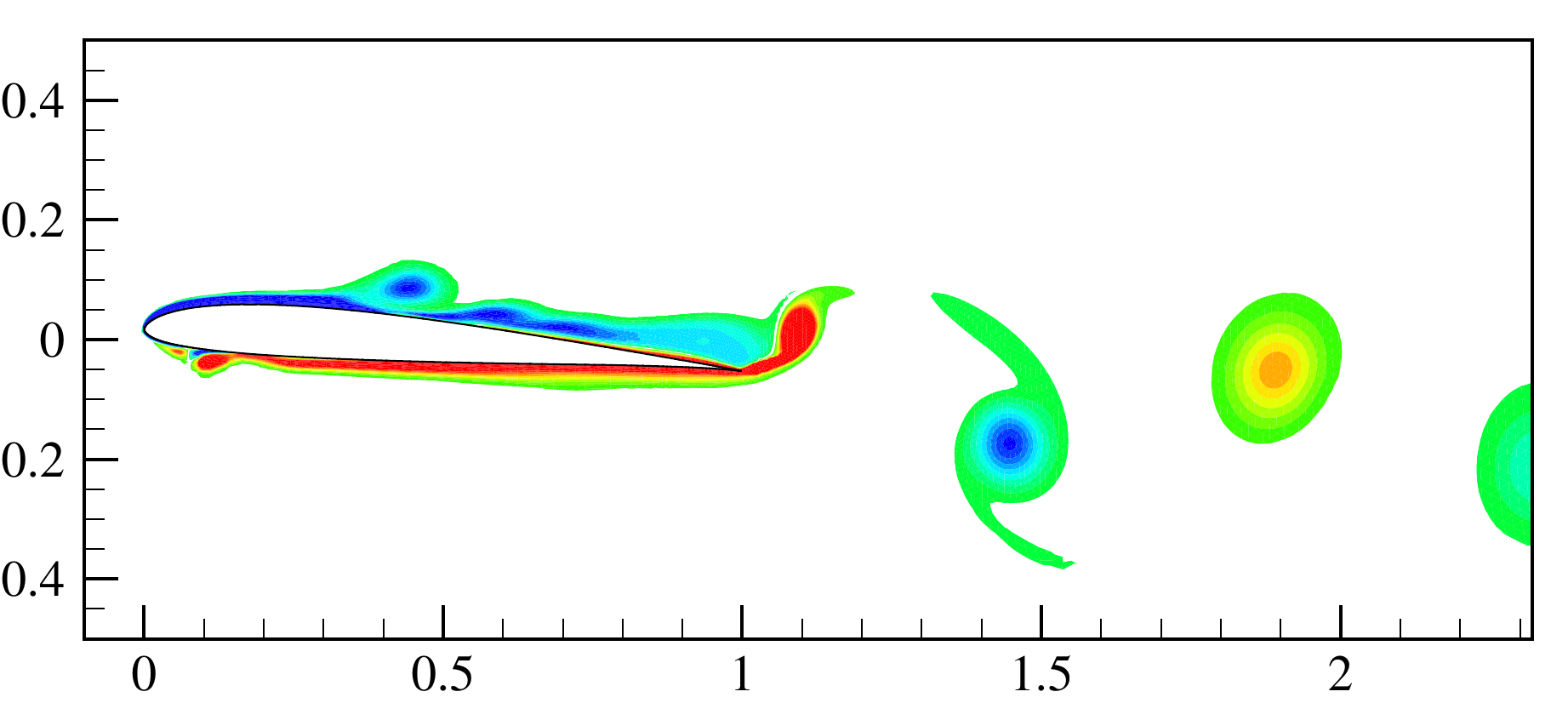}
    \end{minipage}  
  }
  \\
  \subfloat[rDG-ALE, Ma = 0.05]{
      \hspace{0.7in}
      \includegraphics[trim= 2cm 1.5cm 2.2cm 10.5cm,clip,width = 3.1in]
      {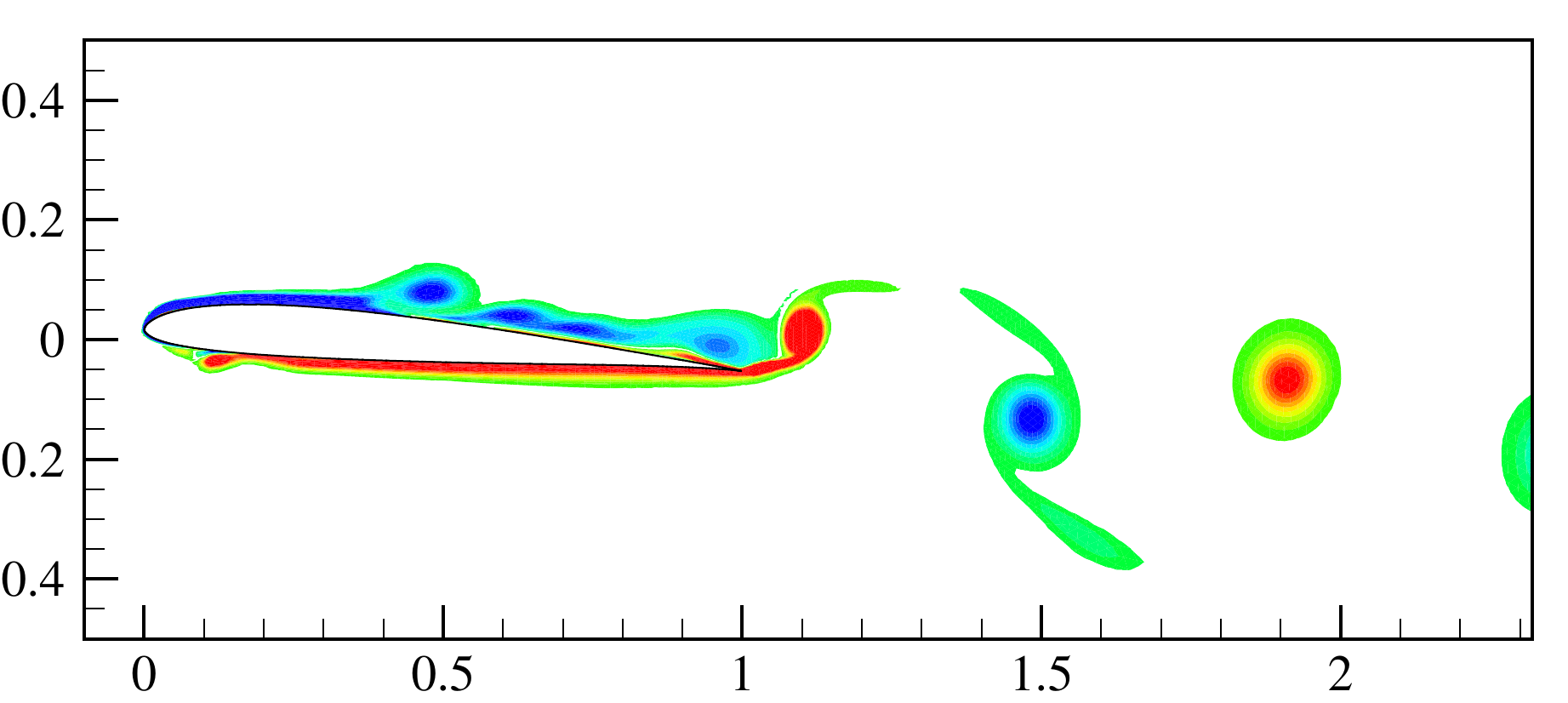}
  }
  \caption{Comparison of normalized vorticities with experiment at $t=2/4 T$  for the plunging SD7003 case.}
  \label{fig:nsSD7003:vorticity:0.5T}
\end{figure}

\section{Conclusions}
\label{sec:conclusion}
An arbitrary Lagrangian-Eulerian (ALE) formulation in the context of 
reconstructed discontinuous Galerkin (rDG) method is proposed and 
presented, for the compressible flows over domains on moving and 
deforming grids with curved elements. 
The Taylor basis functions for the rDG method are defined on the time-
dependent physical element, on which also the integration and 
computations are performed.
A third order ESDIRK3 temporal scheme is responsible for time marching. 
The GCL condition is ensured by modifying the grid velocity terms on 
the right-hand side of the discretized equations.
The radial basis function (RBF) interpolation method is used to 
provide the mesh motion for the interior nodes, given the motion of the boundary nodes.
The numerical results show that the designed spatial and temporal orders of accuracy are achieved.
The simulated results for the moving airfoils problems are compared 
with the experimental or numerical data in the literature, showing the 
capability of this rDG-ALE method for solving moving or deforming domain problems.

\section*{References}

\bibliography{mybibfile}

\end{document}